\documentclass[11pt,a4paper]{article}
\usepackage{jheppub}
\usepackage{graphicx}
\usepackage{amsmath}
\usepackage{bm}
\usepackage{bbm}
\title{Higher-order relativistic corrections to gluon fragmentation into
spin-triplet $\bm{S}$-wave quarkonium}
\author[a]{Geoffrey T. Bodwin,}
\author[b]{U-Rae Kim}
\author[b]{and Jungil Lee}
\affiliation[a]{HEP Division,
Argonne National Laboratory, \\
9700 South Cass Avenue, Lemont, IL 60439, U.S.A.}
\affiliation[b]{Department of Physics, Korea University,
\\ Seoul 136-713, Korea.}
\emailAdd{gtb@hep.anl.gov}
\emailAdd{sadafada@korea.ac.kr}
\emailAdd{jungil@korea.ac.kr}
\abstract{We compute the relative-order-$v^4$ contribution to gluon
fragmentation into quarkonium in the ${}^3S_1$ color-singlet channel,
using the nonrelativistic QCD (NRQCD) factorization approach. The QCD
fragmentation process contains infrared divergences that produce single
and double poles in $\epsilon$ in $4-2\epsilon$ dimensions. We devise
subtractions that isolate the pole contributions, which ultimately are
absorbed into long-distance NRQCD matrix elements in the NRQCD matching
procedure. The matching procedure involves two-loop renormalizations of
the NRQCD operators. The subtractions are integrated over the phase space
analytically in $4-2\epsilon$ dimensions, and the remainder is
integrated over the phase-space numerically. We find that the order-$v^4$
contribution is enhanced relative to the order-$v^0$ contribution. 
However, the order-$v^4$ contribution is not important numerically at 
the current level of precision of quarkonium-hadroproduction phenomenology.
We also estimate the contribution to hadroproduction from gluon fragmentation
into quarkonium in the ${}^3P_J$ color-octet channel and find that it is 
significant in comparison to the complete next-to-leading-order-in-$\alpha_s$
contribution in that channel.}
\arxivnumber{1208.5301}
\begin{document}
\preprint{ANL-HEP-PR-12-65}
\newcommand{\tensor}[1]{\stackrel{\leftrightarrow}{#1}}
\newcommand{\tenD}{-\tfrac{i}{2}\!\!\stackrel{\leftrightarrow}{\bm{D}}}
\maketitle
%%%%%%%%%%%%%%%%%%%%%%%%%%%%%%%%%%%%%%%%%%%%%%%%%%%%%%%%%%%%%%%%%%%%%%
%%%%%%%%%%%%%%%%%%%%%%%%%%%%%%%%%%%%%%%%%%%%%%%%%%%%%%%%%%%%%%%%%%%%%%
\section{Introduction\label{intro}}
%%%%%%%%%%%%%%%%%%%%%%%%%%%%%%%%%%%%%%%%%%%%%%%%%%%%%%%%%%%%%%%%%%%%%%
%%%%%%%%%%%%%%%%%%%%%%%%%%%%%%%%%%%%%%%%%%%%%%%%%%%%%%%%%%%%%%%%%%%%%%
Heavy-quarkonium production in hard-scattering collisions has a long and
rich history of experimental measurements and theoretical calculations
\cite{Brambilla:2010cs}. Intense efforts in this area are expected to
continue as the Large Hadron Collider (LHC) makes available data with
unprecedented momentum transfers and statistics. 

In recent years, a great deal of theoretical effort has been focused on
the nonrelativistic QCD (NRQCD) factorization approach \cite{BBL} 
to calculations of quarkonium production rates. In this approach, it is
conjectured that the inclusive quarkonium production cross section
at large transverse momentum (${{p}_{}}_{T}$) can be written as a
sum of products of short-distance coefficients and long-distance matrix
elements (LDMEs):
%--------------------------------------
\begin{equation}                                                       
\sigma(H)=\sum_n F_n(\mu_{\Lambda})
\langle 0|{\cal O}_n^H(\mu_{\Lambda})|0\rangle.                        
\label{NRQCD-fact}                                            
\end{equation} 
%--------------------------------------
Here, $\mu_{\Lambda}$ is the factorization scale, which is the cutoff
of the effective field theory NRQCD. A short-distance coefficient
$F_n(\mu_{\Lambda})$ is, in essence, the partonic cross section to
produce a heavy-quark-antiquark ($Q\bar Q$) pair with certain quantum
numbers, convolved with parton distribution functions. The
short-distance coefficients can be calculated as perturbation series
in the strong-coupling constant $\alpha_s$. A production LDME 
${\langle 0|\cal O}_n^H(\mu_{\Lambda})|0\rangle$ 
is the probability for a $Q\bar Q$ pair with certain quantum numbers
to evolve into a particular heavy-quarkonium state. It is expressed
as the vacuum expectation value of a four-fermion operator
%--------------------------------------
\begin{equation}                                             
{\cal O}_n^H(\mu_{\Lambda})=\langle 0|\chi^\dagger \kappa_n\psi
{\cal P}_{H(P)} \psi^\dagger  
\kappa'_n\chi|0\rangle,
\end{equation}
%--------------------------------------
where $\psi^\dagger$ and $\chi$ are two-component (Pauli) fields that
create a heavy quark and a heavy antiquark, respectively, and $\kappa_n$
and $\kappa_n'$ are combinations of Pauli and color matrices.\footnote{
It was pointed out by Nayak, Qiu, and Sterman that gauge invariance
requires that the definitions of the NRQCD LDMEs include
Wilson lines that run from the quark and antiquark fields to infinity
\cite{Nayak:2005rw,Nayak:2005rt}. For simplicity, we have omitted these
Wilson lines here.}
%--------------------------------------
\begin{equation}
{\cal P}_{H(P)}=\sum_X|H(P)+X\rangle\langle H(P)+X|
\label{projection}
\end{equation}
%--------------------------------------
is a projection onto a state consisting of a quarkonium $H$, with
four-momentum $P$, plus anything. ${\cal P}_{H(P)}$ contains a sum over
any quarkonium polarization quantum numbers that are not specified
explicitly. The NRQCD LDMEs are evaluated in the rest frame of the
quarkonium, in which $P=(M,\bm{0})$, where $M$ is the quarkonium mass.
In the remainder of this paper, we suppress the momentum argument of
${\cal P}_{H(P)}$ in NRQCD LDMEs.

The production LDMEs for the evolution of color-singlet $Q\bar Q$ pairs
into a quarkonium state are related to the color-singlet quarkonium
decay LDMEs. These color-singlet production LDMEs can be determined
from comparison of theory with quarkonium production or decay data or
from lattice QCD calculations. However, the production LDMEs for the
evolution of color-octet $Q\bar Q$ pairs into a quarkonium state can
be determined, at least at present, only through comparison of theory
with experimental quarkonium-production data.

Complete calculations of short-distance coefficients in the NRQCD
factorization approach now exist through next-to-leading order (NLO) in
$\alpha_s$ for production of the $J/\psi$ and the $\psi(2S)$ in $e^+e^-$
collisions, in $ep$ collisions, and in $p\bar p$ and $pp$ collisions
\cite{Brambilla:2010cs,Butenschoen:2009zy,Ma:2010yw,Butenschoen:2010rq,%
Butenschoen:2010px,Ma:2010jj,Butenschoen:2011yh,Butenschoen:2011ks,%
Butenschoen:2012px,Chao:2012iv,Butenschoen:2012qh,Gong:2012ug}. 
These calculations include the
contributions from all of the color-octet channels through relative
order $v^4$, as well as the contribution of the color-singlet channel 
at leading order (LO) in $v$. Specifically, the calculations include the
contributions of the ${}^3S_1$, $^1S_0$ and ${}^3P_J$ color-octet
channels and the ${}^3S_1$ color-singlet channel at the leading 
nontrivial order in $v$ in each channel. Here, $v$ is half the
relative velocity of the heavy quark and the heavy antiquark in the
quarkonium rest frame. $v^2\approx 0.22$ for the $J/\psi$, and
$v^2\approx 0.1$ for the $\Upsilon$. These theoretical results are 
generally compatible with experimental measurements of quarkonium 
production cross sections. However, significant discrepancies remain 
between theoretical predictions for quarkonium polarization and 
experimental measurements \cite{Chao:2012iv,Butenschoen:2012qh,Gong:2012ug}. 
These discrepancies might point to as-yet-uncalculated theoretical 
contributions, to experimental difficulties, to a failure of convergence 
of the NRQCD series in $\alpha_s$ or $v$, or to a failure of the 
NRQCD factorization conjecture itself.

The calculations at NLO in $\alpha_s$ have revealed very large
corrections in that order to the ${}^3P_J$ color-octet channel and
the ${}^3S_1$ color-singlet channel. Large corrections have also been
found in a calculation of the real-emission contributions to $\Upsilon$
hadroproduction at next-to-next-to-leading order in $\alpha_s$
\cite{Artoisenet:2008fc}. The large corrections are the result of
kinematic enhancements of the higher-order cross sections at
high ${{p}_{}}_{T}$ relative to the LO cross sections. The sizes of these corrections 
have cast some doubt on the convergence of the
perturbation series. 

It has been suggested recently that the large higher-order corrections
to quarkonium production can be brought under control by re-organizing
the perturbation series according to the ${{p}_{}}_{T}$ behavior of the
various contributions \cite{Kang:2011zza}. In this approach, the cross
section can be shown to factorize into convolutions of hard-scattering
cross sections with fragmentation functions. The factorization holds up
to corrections of relative order $m_c^4/{{p}_{}}_{T}^4$, where $m_c$
is the charm-quark mass. The factorized
cross section consists of a leading contribution, which arises from
single-particle fragmentation into a quarkonium and falls as
$1/{{p}_{}}_{T}^4$ in the partonic cross section and a first subleading
contribution, which arises from two-particle fragmentation into a
quarkonium and falls as $1/{{p}_{}}_{T}^6$ in the partonic cross
section. If NRQCD factorization holds, then the various fragmentation
functions can be expressed in terms of a sum of products of
short-distance coefficients and NRQCD LDMEs. This picture has been shown
to account for the large corrections at NLO in $\alpha_s$ in the
${}^3S_1$ color-singlet channel \cite{Kang:2011mg}.

In this paper, we compute the NRQCD short-distance coefficient for gluon
fragmentation into a ${}^3S_1$ color-singlet $Q\bar Q$ pair in relative
order $v^4$. The short-distance coefficients for gluon fragmentation in
this channel have been computed in relative order $v^0$
\cite{Braaten:1993rw,Braaten:1995cj} and relative order $v^2$
\cite{Bodwin:2003wh}. In both cases, the contributions are not important
phenomenologically. Nevertheless, it is worthwhile to consider the
order-$v^4$ contribution for two reasons. First, this contribution is
interesting theoretically because it is in order $v^4$ that the ${}^3S_1$
color-singlet fragmentation channel first develops soft divergences in
full QCD. As we shall see, these soft divergences in full QCD correspond
to soft divergences in the LDMEs for the ${}^3S_1$ and ${}^3P_J$ 
color-octet channels and cancel in the short-distance coefficients,
as is required by NRQCD factorization. A second motivation for examining
the order-$v^4$ contribution is that it is potentially large.
Contributions from gluon fragmentation into the ${}^3S_1$ and ${}^3P_J$
color-octet channels are known to be important phenomenologically. The
${}^3S_1$ color-singlet channel mixes with these channels in order
$v^4$, and the partitioning of the various contributions is controlled
by single and double logarithms of the factorization scale.
Therefore, it is plausible that the order-$v^4$ contributions to the
color-singlet channel could be large.

Our method of calculation is based on the Collins-Soper
definition \cite{Collins:1981uw} of the fragmentation function for a
gluon fragmenting into a quarkonium. We assume that NRQCD factorization
holds, that is, that the fragmentation function can be decomposed into a
sum of products of short-distance coefficients and NRQCD LDMEs. We then
compute the full-QCD fragmentation functions for a gluon fragmenting
into free $Q\bar Q$ states with various quantum numbers. The ultimate
aim is to match these full-QCD fragmentation functions to the
corresponding NRQCD fragmentation functions in order to determine the
NRQCD short-distance coefficients. Some additional details of this
approach can be found in Ref.~\cite{Bodwin:2003wh}. 

This paper is organized as follows. We give the Collins-Soper definition
of the fragmentation function in Sec.~\ref{sec:fragmentation}. 
Section~\ref{sec:factorization} contains the NRQCD factorization formula
for the fragmentation function and also contains a discussion of
the NRQCD LDMEs and short-distance coefficients that are relevant through
relative order $v^4$. The kinematics and variables that we use in our 
calculation are described in Sec.~\ref{sec:kinematics}. 
In Sec.~\ref{sec:full-qcd}, we discuss the calculation of the fragmentation
processes in full QCD. As we have mentioned, an important feature of the
present calculation is that soft divergences arise in the ${}^3S_1$ 
color-singlet channel in both full QCD and NRQCD.
These divergences ultimately cancel in the short-distance coefficients
when we carry out the matching between full QCD and NRQCD. In both full
QCD and NRQCD, we regulate the divergences dimensionally. In the
case of full QCD, we devise subtractions that remove the divergent terms
from the integrand, and we compute the subtraction contributions
analytically. This computation is described in Sec.~\ref{sec:singlet-3s1}.
After we remove the subtraction terms, we calculate the remainder of
the full-QCD contribution in four dimensions, carrying out the integration
numerically. We compute the relevant NRQCD LDMEs for free $Q\bar Q$ states
analytically in dimensional regularization. These calculations are 
described in Sec.~\ref{sec:LDMEs}. We also determine the evolution
equations for the LDMEs and find a discrepancy with a result in 
Ref.~\cite{Gremm:1997dq}. In Sec.~\ref{sec:matching}, we match the NRQCD
and full-QCD fragmentation functions to obtain the short-distance
coefficients, and we present numerical results for them in 
Sec.~\ref{sec:numerical}. Finally, in Sec.~\ref{sec:discussion}, 
we summarize our results.
%%%%%%%%%%%%%%%%%%%%%%%%%%%%%%%%%%%%%%%%%%%%%%%%%%%%%%%%%%%%%%%%%%%%%%
%%%%%%%%%%%%%%%%%%%%%%%%%%%%%%%%%%%%%%%%%%%%%%%%%%%%%%%%%%%%%%%%%%%%%%
\section{Collins-Soper definition of the fragmentation function 
\label{sec:fragmentation}}
%%%%%%%%%%%%%%%%%%%%%%%%%%%%%%%%%%%%%%%%%%%%%%%%%%%%%%%%%%%%%%%%%%%%%%
%%%%%%%%%%%%%%%%%%%%%%%%%%%%%%%%%%%%%%%%%%%%%%%%%%%%%%%%%%%%%%%%%%%%%%
Here, and throughout this paper, we use the following light-cone 
coordinates for a four-vector $V$: 
%--------------------------------------
\begin{subequations}
\begin{eqnarray}
V&=&(V^+,V^-,\bm{V}_\bot)=(V^+,V^-,V^1,V^2),\\
V^+&=&(V^0+V^3)/\sqrt{2},\\
V^-&=&(V^0-V^3)/\sqrt{2}.
\end{eqnarray}
\end{subequations}
%--------------------------------------
The scalar product of two four-vectors $V$ and $W$ is then 
%--------------------------------------
\begin{equation}
V\cdot W=V^+W^-+V^-W^+-\bm{V}_\bot\cdot\bm{W}_\bot.
\end{equation}
%--------------------------------------
The Collins-Soper definition for the fragmentation function for a gluon 
fragmenting into a hadron (quarkonium) $H$ \cite{Collins:1981uw} is 
%--------------------------------------
\begin{eqnarray}
D[g \to H](z,\mu_\Lambda) &=&
\frac{-g_{\mu \nu}z^{d-3} }{ 2\pi k^+ (N_c^2-1)(d-2) }
\int_{-\infty}^{+\infty} dx^- e^{-i k^+ x^-}
\nonumber\\
&&\times
\langle 0 | G^{+\mu}_c(0)
\mathcal{E}^\dagger(0,0,\bm{0}_\perp)_{cb} \; 
\mathcal{P}_{H(P)} \;
\mathcal{E}(0,x^-,\bm{0}_\perp)_{ba} 
G^{+ \nu}_a(0,x^-,\bm{0}_\perp) | 0 \rangle \, .
\nonumber
\\
\label{eq:D-def}
\end{eqnarray}
%--------------------------------------
Here, $z$ is the fraction of the gluon's $+$ component of momentum that
is carried by the hadron, $G_{\mu\nu}$ is the gluon field-strength
operator, $k$ is the momentum of the field-strength operator,
$\mu_\Lambda$ is the factorization scale, and $d=4-2\epsilon$ is the
number of space-time dimensions. There is an implicit average over the
color and polarization states of the initial gluon. The projection
$\mathcal{P}_{H(P)}$ is given in Eq.~(\ref{projection}). The
fragmentation function is evaluated in the frame in which the hadron has
zero transverse momentum: $P=[z k^+,M^2/(2zk^+),\bm{0}_\perp]$. The
operator $\mathcal{E}(0,x^-,\bm{0}_\perp)$ is a path-ordered exponential
of the gluon field:
%--------------------------------------
\begin{equation}
\mathcal{E}(0,x^-,\bm{0}_\perp)_{ba} \;=\; \textrm{P} \exp
\left[ +i g_s\int_{x^-}^\infty dz^- A^+(0,z^-,\bm{0}_\perp) 
\right]_{ba},
\label{eq:E}
\end{equation}
%--------------------------------------
where $g_s=\sqrt{4\pi\alpha_s}$ is the QCD coupling constant and
$A^\mu(x)$ is the gluon field. Both $A_\mu$ and $G_{\mu\nu}$ are SU(3)
matrices in the adjoint representation. The expression (\ref{eq:D-def})
is manifestly gauge invariant. We use the Feynman gauge in our
calculation.

\begin{figure}%[ht]
\begin{center}
\includegraphics[width=4cm]{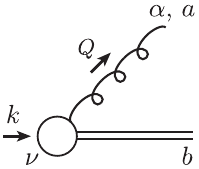}
\end{center}
\vspace{-.5cm}
\caption{Feynman diagram for the vertex that creates a gluon and an
eikonal line. The circle represents the operator $G_{a}^{+\nu}$,
which creates a gluon with momentum $Q$ and polarization and color
indices $\alpha$ and $a$, respectively. $b$ is the color index for the
eikonal line. The operator momentum $k$ is the sum of $Q$ and the
momentum
of the eikonal line.
\label{fig:feynman}}
\end{figure}

The Feynman rules for the perturbative expansion of Eq.~(\ref{eq:D-def})
are given in Ref.~\cite{Collins:1981uw}. The quantity 
$\mathcal{E}(0,x^-,\bm{0}_\perp)$
appears in the Feynman rules as an eikonal line. Owing to the
charge-conjugation properties of the $Q\bar Q$ states that we consider
and the Landau-Yang theorem \cite{landau-thm,Yang:1950rg}, gluon
attachments to the eikonal lines from $\mathcal{E}(0,x^-,\bm{0}_\perp)$ 
do not appear in our calculation. Hence, we
need only the standard QCD Feynman rules, an overall factor
%--------------------------------------
\begin{equation}
\label{eq:c-frag}
C_{\rm frag}=\frac{z^{d-3}k^+}{2\pi(N_c^2-1)(d-2)}
\end{equation}
%--------------------------------------
from Eq.~(\ref{eq:D-def}), and the special Feynman rule for the vertex
that creates a gluon and an eikonal line. That vertex is shown in
Fig.~\ref{fig:feynman}. Its Feynman rule, in momentum space, is a
factor
%--------------------------------------
\begin{equation}
+i\left(g^{\nu\alpha}-\frac{Q^\nu n^\alpha}{k^+}\right)
\delta_{ab},
\label{eq:gL}
\end{equation}
%--------------------------------------
where $k$ is the sum of the momenta of the gluon and the eikonal 
line, $Q$ is the momentum of the gluon, $\alpha$ is the polarization
index of the gluon, and $a$ and $b$ are the color indices, 
respectively, of the gluon and the eikonal line. In the absence of 
interactions with the eikonal lines, $k=Q$. $n$ is a light-like vector
whose components are given by $n=(0,1,\bm{0}_\perp)$.
%==============================================================
%==============================================================

The final-state phase space that is implied by Eq.~(\ref{eq:D-def}) is
%--------------------------------------
\begin{equation}
d\Phi_n=\frac{4\pi M }{S}\;
\delta\left(k^+-P^+-\sum_{i=1}^{n} k_i^+\right)
\theta(k^+)
\prod_{i=1}^n\frac{dk_i^+}{4\pi k^+_{i}}\,
\frac{d^{d-2}\bm{k}_{i\perp}}{(2\pi)^{d-2}}\,
\theta(k_i^+),
\label{eq:phase-space-n}
\end{equation}
%--------------------------------------
where $S$ is the statistical factor for identical particles in the final
state, $k_i$ is the momentum of the $i$th final-state particle, and the
product is over all of the final-state particles except $H$. We use 
nonrelativistic normalization for the state $H$, and so a factor $2M$ 
appears in the phase space in order to cancel the relativistic 
normalization of $H$ in the definition (\ref{eq:D-def}). We use 
relativistic normalization for all particles other than $H$.
%%%%%%%%%%%%%%%%%%%%%%%%%%%%%%%%%%%%%%%%%%%%%%%%%%%%%%%%%%%%%%%%%%%%%%
%%%%%%%%%%%%%%%%%%%%%%%%%%%%%%%%%%%%%%%%%%%%%%%%%%%%%%%%%%%%%%%%%%%%%%
\section{NRQCD factorization\label{sec:factorization}}
%%%%%%%%%%%%%%%%%%%%%%%%%%%%%%%%%%%%%%%%%%%%%%%%%%%%%%%%%%%%%%%%%%%%%%
%%%%%%%%%%%%%%%%%%%%%%%%%%%%%%%%%%%%%%%%%%%%%%%%%%%%%%%%%%%%%%%%%%%%%%
We assume that the fragmentation function for a gluon fragmenting into a 
quarkonium $H$ satisfies NRQCD factorization. Then, in analogy with 
Eq.~(\ref{NRQCD-fact}), we have
%--------------------------------------
\begin{equation}
D[g \to H](z)=
\sum_{n} 
d_{n}(z) \langle 0| \mathcal{O}_{n}^H |0\rangle,
\label{eq:NRQCD-fact-frag}
\end{equation}
%--------------------------------------
where the $\langle 0|\mathcal{O}_{n}^H |0\rangle$ are NRQCD LDMEs and
the $d_{n}(z)$ are the fragmentation short-distance coefficients. We
have suppressed the dependences of $d_{n}(z)$ and  $\langle 0|%
\mathcal{O}_{n}^H |0\rangle$ on the factorization scale $\mu_\Lambda$.
In discussing specific cases, we use the notation
$\langle0|\mathcal{O}_{n_v,n'}^H({}^{2s+1}l_{j}^{[c]})|0\rangle$ for the
LDMEs and the notation $d_{n_v,n'}({}^{2s+1}l_{j}^{[c]})(z)$ for the
short-distance coefficients, where ${}^{2s+1}l_{j}$ is the standard
spectroscopic notation for the angular-momentum quantum numbers of the
corresponding NRQCD operator, and $c$ is the color quantum number of the
NRQCD operator ($1$ or $8$). $n_v$ is the order in $v$, relative to the
leading order, of the field operators and derivatives in
$\mathcal{O}_{n_v,n'}^H$, excluding factors of $v$ from the projection
onto the final state $H$. $n'$ is an integer that is used to distinguish
operators that have the same quantum numbers and order in $v$. We 
denote the contribution of order $v^k$ to $D[g\to H]$ by $D_k[g\to H]$.

We can write 
the fragmentation functions for gluon fragmentation into free 
$Q\bar Q$ states as
%--------------------------------------
\begin{equation}
D[g \to Q\bar Q](z)=
\sum_{n} 
d_{n}(z) \langle 0| \mathcal{O}_{n}^{Q\bar Q} |0\rangle.
\label{eq:NRQCD-fact-frag-free}
\end{equation}
%--------------------------------------
Since the short-distance coefficients $d_{n}(z)$ are independent of the
specifics of the hadronic states, the $d_{n}(z)$ in
Eq.~(\ref{eq:NRQCD-fact-frag-free}) are identical to the $d_{n}(z)$ in
Eq.~(\ref{eq:NRQCD-fact-frag}). We determine the $d_{n}(z)$ by computing
the left side of Eq.~(\ref{eq:NRQCD-fact-frag-free}) in full QCD and
comparing it with the right side, in which the free $Q\bar Q$ LDMEs are
computed in NRQCD. Since we choose a factorization scale $\mu_{\Lambda}$
of order the heavy-quark mass $m$, we can carry out this computation in
perturbation theory.
We will denote the contribution of order
$v^k$ to $D[g\to Q\bar Q]$ by $D_k[g\to Q\bar Q]$.

If $H$ is a ${}^3S_1$ quarkonium  state, such as the $J/\psi$, then,
in LO in $v$, we must consider the LDME
%--------------------------------------
\begin{equation}
\label{v0-ops}
\langle 0|\mathcal{O}_{0}^H({}^3S_1^{[1]})|0\rangle=
\langle 0|\chi^\dagger \sigma^i\psi 
\; {\cal P}_{H} \;
\psi^\dagger \sigma^i \chi|0\rangle.
\end{equation}
%--------------------------------------

In relative order $v^2$, we must consider the LDME
%--------------------------------------
\begin{equation}   
\label{v2-ops}                                                   
\langle 0|\mathcal{O}_{2}^H({}^3S_1^{[1]}) |0\rangle =
\frac{1}{2}\langle 0|\chi^\dagger \sigma^i 
(\tenD)^2\psi \; 
{\cal P}_{H} \;
        \psi^\dagger \sigma^i \chi +\textrm{H.~c.} |0\rangle.
\end{equation}
%--------------------------------------

In relative order $v^3$, we must consider the LDME
%--------------------------------------
\begin{equation}
\langle 0|{\cal O}_{0}^H({}^1S_0^{[8]})|0\rangle
=\langle 0|\chi^\dagger T^a\psi\;{\cal P}_{H}\;
\psi^\dagger T^a \chi |0\rangle.
\end{equation}
%--------------------------------------

In relative order $v^4$, we must consider the LDMEs
%--------------------------------------
\begin{subequations}
\label{v4-ops}
\begin{eqnarray}
\langle 0|{\cal O}_{4,1}^H({}^3S_1^{[1]})|0\rangle
&=&                                   
\langle 0|
\chi^\dagger \sigma^i
(\tenD)^2                                                 
\psi
\;{\cal P}_{H}\;
\psi^\dagger \sigma^i
(\tenD)^2 
\chi
|0\rangle,
\\
\langle 0|\mathcal{O}_{4,2}^H({}^3S_1^{[1]}) |0\rangle &=&
\frac{1}{2}\langle 0|\chi^\dagger \sigma^i 
(\tenD)^4
\psi \; 
{\cal P}_{H} \;
        \psi^\dagger \sigma^i \chi +\textrm{H.~c.} |0\rangle,\\
\langle 0|\mathcal{O}_{4,3}^H({}^3S_1^{[1]}) |0\rangle &=&
\frac{1}{2}\langle 0|\chi^\dagger \sigma^i \psi
\;{\cal P}_{H} \;
\psi^\dagger \sigma^i 
(\tensor{\bm{D}}\!\cdot g_s \bm{E}+g_s\bm{E}\cdot\!\tensor{\bm{D}})
\chi\nonumber\\
&&\qquad -\chi^\dagger \sigma^i 
(\tensor{\bm{D}}\!\cdot g_s \bm{E}+g_s\bm{E}\cdot\!\tensor{\bm{D}})\psi
\;{\cal P}_{H} \;
\psi^\dagger \sigma^i \chi |0\rangle,\\
\langle 0| {\cal O}_{0}^H({}^3S_1^{[8]})|0\rangle
&=&                              
\langle 0|
\chi^\dagger \sigma^i T^a
\psi
\;{\cal P}_{H}\;                     
\psi^\dagger \sigma^i T^a
\chi|0\rangle,
\\
\langle 0| \mathcal{O}^{H}_{0} ({}^3P_0^{[1]})
|0 \rangle
&=&
\frac{1}{d-1}
\langle 0|                                      
\chi^\dagger      
(-\tfrac{i}{2}\!\!\stackrel{\leftrightarrow}{\bm{D}}
\cdot \bm{\sigma})
\psi
\;{\cal P}_{H}\;
\psi^\dagger
(-\tfrac{i}{2}\!\!\stackrel{\leftrightarrow}{\bm{D}}
\cdot \bm{\sigma})
\chi
|0\rangle,\phantom{xxx}
\\
\langle 0| \mathcal{O}^{H}_{0} ({}^3P_1^{[1]})
|0 \rangle
&=&
\langle 0|                                      
\chi^\dagger      
(-\tfrac{i}{2}\!\!\stackrel{\leftrightarrow}D
\phantom{}\!\!^{[i}\sigma^{j]})
\psi
\;{\cal P}_{H}\;
\psi^\dagger
(-\tfrac{i}{2}\!\!\stackrel{\leftrightarrow} D
\phantom{}\!\!^{[i}\sigma^{j]})
\chi
|0\rangle,\phantom{xxx}
\\
\langle 0| \mathcal{O}^{H}_{0} ({}^3P_{2}^{[1]})
|0 \rangle
&=&
\langle 0|                                      
\chi^\dagger      
(-\tfrac{i}{2}\!\!\stackrel{\leftrightarrow}{D}
\!\!\phantom{}^{(i}
\sigma^{j)})
\psi
\;{\cal P}_{H}\;
\psi^\dagger
(-\tfrac{i}{2}\!\!\stackrel{\leftrightarrow}{D}
\!\!\phantom{}^{(i}
\sigma^{j)})
\chi
|0\rangle.
\end{eqnarray}
\end{subequations}
%--------------------------------------
Here, the symmetric traceless product is defined by
%--------------------------------------
\begin{equation}
A^{(i}B^{j)}=\frac{1}{2}(A^i B^j+A^j B^i)
-\frac{1}{d-1}\,\delta^{ij}A^k B^k,
\end{equation}
%--------------------------------------
and the antisymmetric product is defined by
%--------------------------------------
\begin{equation}
 A^{[i}B^{j]}=\frac{1}{2}(A^iB^j-A^jB^i).
\end{equation}
%--------------------------------------
For purposes of our calculation, it is also useful to define
%--------------------------------------
\begin{eqnarray}
\langle 0| {\cal O}_{0}^H({}^3P^{[8]})|0\rangle
&=&                                   
\sum_{J=0,\,1,\,2}\langle 0|{\cal O}_{0}^H({}^3P_J^{[8]})
|0\rangle\nonumber\\
&=&\langle 0|                                      
\chi^\dagger      
(-\tfrac{i}{2}\!\!\stackrel{\leftrightarrow}{\bm{D}}
)^r \sigma^n T^a
\psi
\;{\cal P}_{H}\;
\psi^\dagger
(-\tfrac{i}{2}\!\!\stackrel{\leftrightarrow}{\bm{D}})^r \sigma^n T^a
\chi
|0\rangle.
\end{eqnarray}
%--------------------------------------
It was shown in Ref.~\cite{Bodwin:2002hg} that, by making use of the 
NRQCD equations of motion, one can express the LDME $\langle 
0|\mathcal{O}_{4,3}^H({}^3S_1^{[1]}) |0\rangle$ in terms of the LDMEs
$\langle 0|\mathcal{O}_{4,1}^H({}^3S_1^{[1]}) |0\rangle$ and 
$\langle 0|\mathcal{O}_{4,2}^H({}^3S_1^{[1]}) |0\rangle$. (Equivalently, 
one can eliminate the LDME $\langle 0|\mathcal{O}_{4,3}^H({}^3S_1^{[1]}) 
|0\rangle$ by making use of a field redefinition 
\cite{Brambilla:2008zg}.) Hence, we need not consider $\langle 
0|\mathcal{O}_{4,3}^H({}^3S_1^{[1]})
|0\rangle$ in our analysis. 

In the LDMEs $\langle 0|\mathcal{O}_{4,1}^H({}^3S_1^{[1]})
|0\rangle$ and $\langle 0|\mathcal{O}_{4,2}^H({}^3S_1^{[1]}) |0\rangle$,
one can replace $|H+X\rangle\langle H+X|$ in the projector ${\cal
P}_{H(M,\bm{0})}$ with $|H\rangle\langle H|$ (vacuum-saturation
approximation), making an error of relative order $v^4$. If one takes
this approximation and evaluates the LDMEs in dimensional regularization
in a potential model, then they are equal \cite{Bodwin:2006dn}. Since
the static potential model is valid up to corrections of order $v^2$, 
we have
%--------------------------------------
\begin{equation}
\langle
0|\mathcal{O}_{4,1}^H({}^3S_1^{[1]}) |0\rangle =
\langle 0|\mathcal{O}_{4,2}^H({}^3S_1^{[1]}) |0\rangle +{\cal O}(v^2).
\end{equation}
%--------------------------------------
Hence, up to corrections of relative order $v^2$, only the sum of 
short-distance    
coefficients $d_{4,1}({}^3S_1^{[1]})(z)+ d_{4,2}({}^3S_1^{[1]})(z)$ 
appears in the fragmentation function.
%%%%%%%%%%%%%%%%%%%%%%%%%%%%%%%%%%%%%%%%%%%%%%%%%%%%%%%%%%%%%%%%%%%%%%
\subsection{NRQCD factorization formulas for ${g\to J/\psi}$ 
through order $v^4$}
%%%%%%%%%%%%%%%%%%%%%%%%%%%%%%%%%%%%%%%%%%%%%%%%%%%%%%%%%%%%%%%%%%%%%%
In summary, we have the following NRQCD factorization formulas for 
gluon fragmentation into $J/\psi$ through relative order $v^4$.

In relative order $v^0$ we have
%--------------------------------------
\begin{equation}
\label{D0-J/psi}
D_0[g \to J/\psi]
=
d_0[g \to Q\bar{Q}({}^3S_1^{[1]})]
\langle 0|
\mathcal{O}_{0}^{J/\psi}({}^3S_1^{[1]})
|0\rangle.
\end{equation}
%--------------------------------------
The short-distance coefficient $d_0[g \to Q\bar{Q}({}^3S_1^{[1]})]$
was calculated in Refs.~\cite{Braaten:1993rw,Braaten:1995cj}.

In relative order $v^2$ we have
%--------------------------------------
\begin{equation}
\label{D2-J/psi}
D_2[g \to J/\psi]
=
d_2[g \to Q\bar{Q}({}^3S_{1}^{[1]})]
\langle 0|
\mathcal{O}_{2}^{J/\psi}({}^3S^{[1]})
|0\rangle.
\end{equation}
%--------------------------------------
The short-distance coefficient $d_2[g \to Q\bar{Q}({}^3S_1^{[1]})]$
was calculated in Ref.~\cite{Bodwin:2003wh}.

In relative order $v^3$ we have
%--------------------------------------
\begin{equation}
\label{D3-J/psi}
D_3[g \to J/\psi]
=
d_0[g \to Q\bar{Q}({}^1S_0^{[8]})]
\langle 0|
\mathcal{O}_{0}^{J/\psi}({}^1S_0^{[8]})
|0\rangle.
\end{equation}
%--------------------------------------
The short-distance coefficient 
$d_0[g \to Q\bar{Q}({}^1S_0^{[1]})]$ was
calculated in Ref.~\cite{Braaten:1996rp} and differs from the
short-distance coefficient 
$d_0[g \to Q\bar{Q}({}^1S_0^{[8]})]$ in
Eq.~(\ref{D3-J/psi}) only by a color factor, which we provide in
Sec.~\ref{sec:matching}.

In relative order $v^4$ we have
%--------------------------------------
\begin{eqnarray}
\label{D4-Jpsi}
D_{4}[g \to J/\psi]
&=&
\big\{\,
d_{4,1}[g \to Q\bar{Q}({}^3S_1^{[1]})]
+
d_{4,2}[g \to 
Q\bar{Q}({}^3S_1^{[1]})]\,\big\}
\,
\langle 0|
\mathcal{O}_{4}^{J/\psi}({}^3S_1^{[1]})
|0\rangle
\nonumber\\
&+&
d_{0}[g \to Q\bar{Q}({}^3P^{[8]})]
\langle 0|
\mathcal{O}_{0}^{J/\psi}({}^3P^{[8]})
|0\rangle
\nonumber\\
&+&
d_{0}[g \to Q\bar{Q}({}^3S_1^{[8]})]
\langle 0|
\mathcal{O}_{0}^{J/\psi}({}^3S_1^{[8]})
|0\rangle.
\end{eqnarray}
%--------------------------------------
The short-distance coefficient $d_{0}[g \to Q\bar{Q}({}^3S_1^{[8]})]$
was calculated at LO in $\alpha_s$ in 
Refs.~\cite{Braaten:1996rp,Bodwin:2003wh}
and at NLO in $\alpha_s$ in Refs.~\cite{BL:gfrag-NLO,Lee:2005jw}. We verify 
the LO calculations in Refs.~\cite{Braaten:1996rp,Bodwin:2003wh} 
in the present paper, giving our result in 
Sec.~\ref{sec:matching}. We compute $d_{0}[g \to Q\bar{Q}({}^3P^{[8]})]$
in this paper, giving the result in Sec.~\ref{sec:matching}. The
short-distance coefficient $d_{0}[g \to Q\bar{Q}({}^3P^{[1]})]$ was
calculated in Refs.~\cite{Braaten:1996rp} and differs from the
short-distance coefficient $d_{0}[g \to Q\bar{Q}({}^3P^{[8]})]$ in
Eq.~(\ref{D4-Jpsi}) only by a color factor. The computation of the
combination of short-distance coefficients $d_{4,1}[g \to
Q\bar{Q}({}^3S_1^{[1]})]+d_{4,2}[g \to Q\bar{Q}({}^3S_1^{[1]})]$ is the
main goal of this paper. The  result of that computation is given in
Sec.~\ref{sec:matching}.
%%%%%%%%%%%%%%%%%%%%%%%%%%%%%%%%%%%%%%%%%%%%%%%%%%%%%%%%%%%%%%%%%%%%%%
%%%%%%%%%%%%%%%%%%%%%%%%%%%%%%%%%%%%%%%%%%%%%%%%%%%%%%%%%%%%%%%%%%%%%%
\section{Kinematics\label{sec:kinematics}}
%%%%%%%%%%%%%%%%%%%%%%%%%%%%%%%%%%%%%%%%%%%%%%%%%%%%%%%%%%%%%%%%%%%%%%
%%%%%%%%%%%%%%%%%%%%%%%%%%%%%%%%%%%%%%%%%%%%%%%%%%%%%%%%%%%%%%%%%%%%%%
In the calculations to follow, in both full QCD and NRQCD, we employ the 
following kinematics.  

We take the $Q$ and the $\bar Q$ to be free (on-shell) states with
momenta
%--------------------------------------
\begin{subequations}
\begin{eqnarray}
p&=&\tfrac{1}{2}P+q,\\
\bar{p}&=&\tfrac{1}{2}P-q,
\end{eqnarray}
\end{subequations}
%--------------------------------------
respectively. The heavy quark has
three-momentum $\bm{q}$ in the $Q \bar{Q}$ rest frame, and, so, the
invariant mass of the $Q\bar Q$ state is 
%--------------------------------------
\begin{equation}
P^2 =M^2= 4E^2, 
\end{equation}
%--------------------------------------
where
%--------------------------------------
\begin{equation}
E=\sqrt{m^2+\bm{q}^2}.
\end{equation}
%--------------------------------------

We work in the frame in which the transverse momentum of the $Q\bar Q$
pair vanishes. In this frame, the initial-state gluon, the final-state
$Q\bar Q$ pair and the final-state gluons, respectively, have the 
momenta
%--------------------------------------
\begin{subequations}
\begin{eqnarray}
k&=&\left(k^+,k^-=\frac{k^2+ (P_\perp/z)^2}{2k^+},
-\frac{\bm{P}_\perp}{z}\right),\\
P&=&\left(zk^+,\frac{M^2}{2zk^+},\bm{0}_\perp\right),
\\
k_1&=&\left(z_1k^+,\frac{k_{1\perp}^2}{2z_1k^+},\bm{k}_{1\perp}\right),
\\
k_2&=&\left(z_2k^+,\frac{k_{2\perp}^2}{2z_2k^+},\bm{k}_{2\perp}\right),
\end{eqnarray}
\end{subequations}
%--------------------------------------
where we have introduced the longitudinal momentum fractions
%--------------------------------------
\begin{subequations}
\begin{eqnarray}
z&=&\frac{P^+}{k^+},
\\
z_1&=&\frac{k_1^+}{k^+},
\\
z_2&=&\frac{k_2^+}{k^+}.
\end{eqnarray}
\end{subequations}
%--------------------------------------
Because of the conservation of four-momentum, $k=P+k_1+k_2$, the momenta 
$k$, $k_1$ and $k_2$ depend implicitly on $P$, and, therefore, on $q$. 
We can make the dependence on $q$ explicit by writing quantities in 
terms of dimensionless momenta
%--------------------------------------
\begin{subequations}
\label{dimensionless-mom}
\begin{eqnarray}
\bar{P}&=&
\frac{P}{\sqrt{P^2}},
\\
\bar{k}&=&
\frac{k}{\sqrt{P^2}},
\\
\bar{k}_1&=&
\frac{k_1}{\sqrt{P^2}},
\\
\bar{k}_2&=&
\frac{k_2}{\sqrt{P^2}}.
\end{eqnarray}
\end{subequations}
%--------------------------------------
It is also useful to express the Lorentz invariants in terms of the 
following dimensionless variables:
%--------------------------------------
\begin{subequations}
\begin{eqnarray}
e_1&=&
\bar{k}_1\cdot \bar{P},
\\
e_2&=&
\bar{k}_2\cdot \bar{P},
\\
x&=&
\bar{k}_1\cdot \bar{k}_2
=e_1 e_2(1-\hat{\bm{k}}_1\cdot\hat{\bm{k}}_2),
\end{eqnarray}
\end{subequations}
%--------------------------------------
where $\hat{\bm{k}}_i$ is the unit vector that is parallel to
the three-vector $\bm{k}_i$ in the $Q\bar{Q}$ rest frame.

The phase space in Eq.~(\ref{eq:phase-space-n}) can be expressed
in terms of the dimensionless variables as
%--------------------------------------
\begin{subequations}
\label{dPhin-org}
\begin{eqnarray}
\label{dPhi0-org}
d\Phi_0
&=&
\frac{4\pi M}{k^+} \,\delta(1-z),
\\
\label{dPhi1-org}
d\Phi_1
&=&
\frac{4\pi M^{d-1}}{k^+} \,\theta(z_1)\,\delta(1-z-z_1)\,
\frac{d z_1}{4\pi z_1} \frac{d^{d-2}\bar{\bm{k}}_{1\perp}}{(2\pi)^{d-2}},
\\
\label{dPhi2-org}
d\Phi_2
&=&
\frac{4\pi M^{2d-3}}{Sk^+} 
\,\theta(z_1)\,\theta(z_2)\,\delta(1-z-z_1-z_2)
\,\frac{d z_1}{4\pi z_1}\frac{d z_2}{4\pi z_2} 
\frac{d^{d-2}\bar{\bm{k}}_{1\perp}}{(2\pi)^{d-2}}
\frac{d^{d-2}\bar{\bm{k}}_{2\perp}}{(2\pi)^{d-2}}.
\phantom{xxx}
\end{eqnarray}
\end{subequations}
%--------------------------------------
We have not replaced the overall factor $1/k^+$
in Eq.~(\ref{dPhin-org}) with $1/(\bar{k}^+\sqrt{P^2})$, 
because it ultimately will be
cancelled by the factor $k^+$ in $C_{\rm frag}$ in 
Eq.~(\ref{eq:c-frag}).
The ranges of the variables $z$, $z_1$ and $z_2$ are 
completely determined by the $\delta$ and $\theta$ functions. 
When we expand the fragmentation function in powers of $q$, 
it is convenient to make use of the phase space at LO in $q$,
%--------------------------------------
\begin{equation}
\label{dPhin-tilde}
d\tilde{\Phi}_n=d\Phi_n\big|_{\bm{q}\to\bm{0}},
\end{equation}
%--------------------------------------
where $\bm{q}\to\bm{0}$ means that, in the phase space in
Eq.~(\ref{dPhin-org}), we replace $M$ with $2m$. Then $d\Phi_n$
can be expressed in terms of $d\tilde{\Phi}_n$ as follows:
%--------------------------------------
\begin{subequations}
\label{dPhin-tilde-relation}
\begin{eqnarray}
\label{dPhi0-tilde-relation}
d\Phi_0&=&\frac{E}{m} d\tilde{\Phi}_0,
\\
\label{dPhi1-tilde-relation}
d\Phi_1&=&\left(\frac{E}{m}\right)^{3-2\epsilon}d\tilde{\Phi}_1,
\\
\label{dPhi2-tilde-relation}
d\Phi_2&=&\left(\frac{E}{m}\right)^{5-4\epsilon}d\tilde{\Phi}_2.
\end{eqnarray}
\end{subequations}
%--------------------------------------
We express our results for the fragmentation contributions 
in terms of integrals over the phase spaces $d\tilde{\Phi}_n$: 
%--------------------------------------
\begin{equation}
\label{def:D-tilde}
\tilde{D}[g\to Q\bar{Q}]=D[g\to Q\bar{Q}]
\big|_{d\Phi_n\to d\tilde{\Phi}_n}.
\end{equation}
%--------------------------------------
The factors of $E/m$ in Eq.~(\ref{dPhin-tilde-relation}) are then an 
additional source of relativistic corrections. 
%%%%%%%%%%%%%%%%%%%%%%%%%%%%%%%%%%%%%%%%%%%%%%%%%%%%%%%%%%%%%%%%%%%%%%
%%%%%%%%%%%%%%%%%%%%%%%%%%%%%%%%%%%%%%%%%%%%%%%%%%%%%%%%%%%%%%%%%%%%%%
\section{Full-QCD calculations\label{sec:full-qcd}}
%%%%%%%%%%%%%%%%%%%%%%%%%%%%%%%%%%%%%%%%%%%%%%%%%%%%%%%%%%%%%%%%%%%%%%
%%%%%%%%%%%%%%%%%%%%%%%%%%%%%%%%%%%%%%%%%%%%%%%%%%%%%%%%%%%%%%%%%%%%%%
In this section, we compute the relevant fragmentation functions for
free $Q\bar Q$ states in full QCD. We have carried out the
calculations by writing independent codes using \textsc{reduce}
\cite{REDUCE} and using the \textsc{feyncalc} package \cite{Mertig:an}
in \textsc{mathematica} \cite{MATH}. At each stage of the calculations
we have checked that the independent codes give identical results.

The computations are carried out in $d=4-2\epsilon$ dimensions with
dimensional-regulariza-%
tion scale $\mu$. We use the
modified-minimal-subtraction ($\overline{\rm MS}$) scheme throughout.
Then, in $d=4-2\epsilon$ dimensions, there is a factor 
$[\mu^2\exp({{\gamma}_{}}_{\rm E})/(4\pi)]^\epsilon$ 
that is associated with
each factor of the strong coupling $\alpha_s$, where 
${{\gamma}_{}}_{\rm E}$ is the Euler-Mascheroni constant.

In computing the $Q\bar Q$ fragmentation functions, it is convenient to
make use of projection operators for the spin and color states of the
$Q\bar{Q}$ pair. The projection operators for a $Q\bar{Q}$
pair in the color-singlet and color-octet configurations are
%--------------------------------------
\begin{subequations}
\label{eq:PJ-color}
\begin{eqnarray}
\Lambda_{1}&=&\frac{1}{\sqrt{N_c}} 
\mathbbm{1},\label{eq:PJ-color-singlet}\\
\Lambda^a_{8}&=&\sqrt{2} T^a,
\label{eq:PJ-color-octet}
\end{eqnarray}
\end{subequations}
%--------------------------------------
where 
$\mathbbm{1}$ and $T^a$ are the identity matrix and the generator of the
fundamental (triplet) representation of SU(3), $a$ is the
adjoint-representation color index ($a=1,\ldots,N_c^2-1$),  and
$N_c=3$. Spin-projection operators at LO in $v$ were first
given in
Refs.~\cite{Barbieri:1975am,Barbieri:1976fp,Chang:1979nn,%
Guberina:1980dc,Berger:1980ni}.
Projectors accurate to all orders in $v$ were
given in Ref.~\cite{Bodwin:2002hg}. For the spin-triplet state, the
projection operator, correct to all orders in $v$, is 
%--------------------------------------
\begin{equation}
\Lambda(P,q,\epsilon_{S}^*)=
N(\not\!{\overline{p}}-m)\not\!{\epsilon}^*_{S}\,
\frac{\not\!{P}+2E}{4E}
(\not\!{p}+m),
\label{eq:PJ-spin}
\end{equation}
%--------------------------------------
where $\epsilon_{S}$ is the spin polarization of the $Q\bar{Q}$ pair,
and $N=[2\sqrt{2}E(E+m)]^{-1}$. Note that we use nonrelativistic
normalization for the heavy-quark spinors.

The use of the spin projection (\ref{eq:PJ-spin}) in $d$ dimensions
requires some justification. It accounts for only the $d-1$ vector
polarization states, which, in the $Q\bar Q$ rest frame, correspond to
the $d-1$ Pauli matrices $\sigma_i$. In general, in $d$ dimensions, one
must consider states that correspond to products of the $\sigma_i$ that
are linearly independent of the $\sigma_i$ \cite{Braaten:1996rp}.  These
additional states vanish as $\epsilon$ goes to zero. Hence, they can
contribute only in conjunction with a pole in $\epsilon$. The poles in
$\epsilon$ in our calculation correspond to soft divergences. The
divergent parts of soft interactions arise from the convection current
on fermions lines, and, hence, do not change the fermion spin.
Therefore, the additional states that correspond to products of the
$\sigma_i$ never mix in our calculation with the vector states that
correspond to the $\sigma_i$. Consequently, we need consider only the
$d-1$ vector states in our calculation.\footnote{Some elements of this
argument were presented in Ref.~\cite{Petrelli:1997ge}.}

The spin-triplet, color-singlet part of an amplitude $\mathcal{C}$ is
%--------------------------------------
\begin{equation}
\mathcal{M}=
\textrm{Tr}[\,\mathcal{C}(\Lambda\otimes \Lambda_1)\,],
\end{equation} 
%--------------------------------------
and the spin-triplet, color-octet part of an amplitude $\mathcal{C}$ is 
%--------------------------------------
\begin{equation}
\mathcal{M}^a=
\textrm{Tr}[\,\mathcal{C}(\Lambda\otimes \Lambda_8^a)\,],
\end{equation}
%--------------------------------------
where the traces are over the Dirac and color indices.
The amplitude $\mathcal{C}$ includes the propagator of the initial 
gluon, as well as the associated polarization factor in 
Eq.~(\ref{eq:gL}).
In our calculation, the amplitudes $\mathcal{C}$, $\mathcal{M}$, and 
$\mathcal{M}^a$ are all expressed in terms of the dimensionless variables in 
Eq.~(\ref{dimensionless-mom}) or invariants that are formed from them, 
and so the dependence on $q$ is explicit.
                
The $S$-wave part of $\mathcal{M}$ (with color index suppressed in the
color-octet case) can be written as an expansion in powers of
$v^2=\bm{q}^2/m^2$:
%--------------------------------------
\begin{equation}
\mathcal{M}_{S}=
\mathcal{M}_{S0}
+\mathcal{M}_{S2}+
\mathcal{M}_{S4}+O(\bm{q}^6/m^6),
\label{eq:M-S-wave}
\end{equation}
%--------------------------------------
where
%--------------------------------------
\begin{subequations}
\label{eq:M-S-2}
\begin{eqnarray}
\mathcal{M}_{S0}&=&\left(\mathcal{M}\right)_{\bm{q}\to \bm{0}}\;
                       ,\\
\mathcal{M}_{S2}&=&
             \frac{\bm{q}^2}{2!(d-1)}\,I^{\alpha\beta}
                              \left(\frac{\partial^2 \mathcal{M}}
                                {\partial q^\alpha\partial q^\beta}
                              \right)_{\bm{q}\to \bm{0}}\;
                       ,\\
\mathcal{M}_{S4}&=&
             \frac{\bm{q}^4}{4!(d-1)(d+1)}\,
             I^{\alpha\beta\gamma\delta} 
             \left(\frac{\partial^4 \mathcal{M}}
{\partial q^\alpha\partial q^\beta \partial q^\gamma \partial q^\delta}
\right)_{\bm{q}\to \bm{0}}\;,
\end{eqnarray}
\end{subequations}
%--------------------------------------
and 
%--------------------------------------
\begin{subequations}
\begin{eqnarray}
\label{I-ab}
I^{\alpha\beta}&=&
-g^{\alpha\beta}+P^\alpha P^\beta/(4E^2),
\\
I^{\alpha\beta\gamma\delta}
&=&
I^{\alpha\beta}I^{\gamma\delta}+
I^{\alpha\gamma}I^{\beta\delta}+I^{\alpha\delta}I^{\beta\gamma}.
\end{eqnarray}
\end{subequations}
%--------------------------------------

In order to project out the $P$-wave part of the amplitude ${\cal M}$, 
we multiply  ${\cal M}$ by the $P$-wave orbital-angular-momentum state
$-\sqrt{d-1}\,\epsilon_{L}^*\cdot \hat{q}$ 
and average over the direction of $\bm{q}$.\footnote{
In some calculations in 
NRQCD, the $P$-wave orbital-angular-momentum state          
is normalized as $-\epsilon_{L}^*\cdot \hat{q}$.}
Here, $\epsilon_{L}$ is the polarization vector for the 
orbital-angular-momentum state, and
$\hat{q}=(0,\hat{\bm{q}})$ in the rest frame of the
$Q\bar{Q}$ pair. Then, the $P$-wave part of the amplitude is
%--------------------------------------
\begin{equation}
\mathcal{M}_{P}=
\mathcal{M}_{P1}
+O(\bm{q}^3/m^3),
\label{eq:M-P-wave}
\end{equation}
%--------------------------------------
where 
%--------------------------------------
\begin{equation}
\mathcal{M}_{P1}=
-\frac{|\bm{q}|}{\sqrt{d-1}}\,
\epsilon^*_{L\alpha}\,I^{\alpha\beta}
\left(\frac{\partial \mathcal{M}}
{\partial q^\beta}
\right )_{\bm{q}\to \bm{0}}\;.
\end{equation}
%--------------------------------------

We define squared amplitudes for the color-singlet 
and color-octet states as
%--------------------------------------
\begin{subequations}
\label{squared-amp}
\begin{eqnarray}
\mathcal{A}({}^{2s+1}l_j^{[1]})
&=&C_{\rm frag}|{\cal M} ({}^{2s+1}l_j^{[1]})|^2,
\\
\mathcal{A}({}^{2s+1}l_j^{[8]})
&=&C_{\rm frag}\sum_{a}
|{\cal M}^{a}({}^{2s+1}l_j^{[8]})|^2,
\end{eqnarray}
\end{subequations}
%--------------------------------------
where $C_{\rm frag}$ is given in Eq.~(\ref{eq:c-frag}), and it is
implicit that there are sums over the spin and orbital-angular-momentum
polarizations of the $Q\bar{Q}$ states and sums over the polarizations of
the initial and final gluons. Note that
%--------------------------------------
\begin{equation}
\sum_\lambda
\epsilon_{S}^{\alpha*}(\lambda)\epsilon_{S}^{\beta}(\lambda)=
\sum_\lambda
\epsilon_{L}^{\alpha*}(\lambda)\epsilon_{L}^{\beta}(\lambda)=
I^{\alpha\beta}.
\end{equation}
%--------------------------------------
We denote the order-$v^k$ contribution to $\mathcal{A}$ by
$\mathcal{A}_k$.
%%%%%%%%%%%%%%%%%%%%%%%%%%%%%%%%%%%%%%%%%%%%%%%%%%%%%%%%%%%%%%%%%%%%%%
\subsection{$D_{0}[g\to Q\bar Q({}^3S_1^{[8]})]$}
%%%%%%%%%%%%%%%%%%%%%%%%%%%%%%%%%%%%%%%%%%%%%%%%%%%%%%%%%%%%%%%%%%%%%%
\begin{figure}%[ht]
\begin{center}
\includegraphics[height=5cm]{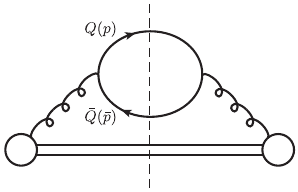}
\end{center}
\vspace{-.5cm}
\caption{Feynman diagram for the fragmentation process
$g\to Q\bar{Q}({}^3S_1^{[8]})$. 
The dashed line represents the final-state cut.
The momenta for $Q$ and $\bar{Q}$ on the left side of the 
cut are 
$p=\tfrac{1}{2}P+q$ and
$\bar{p}=\tfrac{1}{2}P-q$, respectively.
The momenta on the right side of the cut are 
$\tfrac{1}{2}P+q'$ and
$\tfrac{1}{2}P-q'$, respectively. Here,
$|\bm{q}'|=|\bm{q}|$ in the rest frame of the $Q\bar{Q}$
pair, but we distinguish the directions of $\bm{q}$ and $\bm{q}'$ in 
order to be able to project out orbital-angular-momentum states in the 
amplitude and its complex conjugate.
\label{fig:frag1}}
\end{figure}
%==============================================================
The diagram for gluon fragmentation into a ${}^3S_1$ color-octet $Q\bar Q$ 
pair at order $v^0$ and $\alpha_s^1$ is shown in 
Fig.~\ref{fig:frag1}. A 
straightforward computation yields
%--------------------------------------
\begin{equation}
\mathcal{A}_0({}^3S_1^{[8]})
=
\frac{\alpha_s k^+}
{8m^4}\left(
\frac{\mu^2}{4\pi}e^{{{\gamma}_{}}_{\rm E}}
\right)^\epsilon.
\end{equation}
%--------------------------------------
Carrying out the trivial integration over the phase space 
$d\tilde\Phi_0$
in Eqs.~(\ref{dPhi0-org}) and (\ref{dPhi0-tilde-relation}), we obtain 
%--------------------------------------
\begin{equation}
\label{D3s18}
D_0[g\to Q\bar Q({}^3S_1^{[8]})]
=
\frac{\pi\alpha_s}{m^3}\left(
\frac{\mu^2}{4\pi}e^{{{\gamma}_{}}_{\rm E}}
\right)^\epsilon\delta(1-z).
\end{equation}
%%%%%%%%%%%%%%%%%%%%%%%%%%%%%%%%%%%%%%%%%%%%%%%%%%%%%%%%%%%%%%%%%%%%%%
\subsection{$D_2[g\to Q\bar Q({}^3P^{[8]})]$}
%%%%%%%%%%%%%%%%%%%%%%%%%%%%%%%%%%%%%%%%%%%%%%%%%%%%%%%%%%%%%%%%%%%%%%
\begin{figure}%[ht]
\begin{center}
\includegraphics[height=5cm]{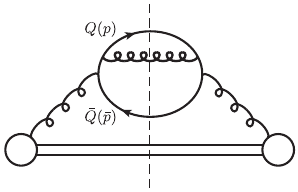}
\end{center}
\vspace{-.5cm}
\caption{
One of the four Feynman diagrams for the fragmentation 
process $g\to Q\bar{Q}({}^3P^{[8]})$. 
Three additional diagrams can be obtained by permuting
the gluon-fermion vertices on the left and right sides of the cut.
\label{fig:frag2}}
\end{figure}
%==============================================================
The diagrams for gluon fragmentation into a ${}^3P^{[8]}$ color-octet 
$Q\bar Q$ pair at LO in $\alpha_s$ and $v$ are shown 
in Fig.~\ref{fig:frag2}. We obtain 
%--------------------------------------
\begin{equation}
\label{msquared-3P8}
\mathcal{A}_1({}^3P^{[8]})
=
\frac{k^+ \pi\alpha_s^2   (N_c^2-4) \bm{q}^2  z^{1-2 \epsilon}}{8N_c(d-1)(d-2) m^8 }
\left(
\frac{\mu^2}{4\pi}e^{{{\gamma}_{}}_{\rm E}}
\right)^{2\epsilon}
\sum_{n=0}^3
\frac{ \rho_n(z) }{e_1^{n} (1+2 e_1)^2},
\end{equation}
%--------------------------------------
where $\rho_n(z)$ are given by
%--------------------------------------
\begin{subequations}
\begin{eqnarray}
\rho_0(z)&=&
(3-4 \epsilon) (2-2 \epsilon-4 z+4 z^2),
\\
\rho_1(z)&=&
2\big[ 5-10 \epsilon+4 \epsilon^2
        -z (5-12 \epsilon)
        +2 z^2 (1-4\epsilon) \big],\\
\rho_2(z)&=&3-12 \epsilon+4 \epsilon^2
    +2 z (3+4 \epsilon)
    -z^2 (5+4 \epsilon),\\
\rho_3(z)&=&-2(1-z)^2.
\end{eqnarray}
\end{subequations}
%--------------------------------------
We carry out the integration over the phase space $d\tilde\Phi_1$ in
Eqs.~(\ref{dPhi1-org}) and (\ref{dPhi1-tilde-relation}) by making
use of the methods that are described in Appendix~\ref{app:integrals}.
Then, we obtain
%--------------------------------------
\begin{subequations}
\label{D3p8}
\begin{eqnarray}
D_{2}[g\to Q\bar Q({}^3P^{[8]})]&=&
\frac{8\alpha_s^2\,\bm{q}^2}{(d-1)m^5}
\,
\frac{N_c^2-4}{4N_c}
(1-\epsilon)
\Gamma(1+\epsilon)
\left(
\frac{\mu^2}{4\pi}e^{{{\gamma}_{}}_{\rm E}}
\right)^{\epsilon}
\left(\frac{\mu^2}{4m^2}e^{{{\gamma}_{}}_{\rm E}}\right)^\epsilon
\nonumber\\
&&\times
\left[-\frac{1}{2\epsilon_{\rm IR}}\, \delta(1-z) 
+  f(z)\right],
\end{eqnarray}
where the finite function $f(z)$ is defined by
\begin{eqnarray}
\label{eq:fz}
f(z)&=&
\left[\frac{1}{(1-z)^{1+2\epsilon}}\right]_+
+\frac{1}{4(1-\epsilon)^{2}}
\Bigg\{
\frac{(1-z)^{-\epsilon}-(1-z)^{-2\epsilon}}{\epsilon}
\,(13-7z)
\nonumber\\
&&
+(1-z)^{-2\epsilon}[ 10+ 3z-5z^2 + 2\epsilon(2-2z+z^2)-4 \epsilon^2   ]
\nonumber\\
&&
+\frac{1}{2}(1-z)^{-\epsilon} [-28+15z      
+  \epsilon(8-11z)   +4 \epsilon^2 z ]\Bigg\}.
\end{eqnarray}
\end{subequations}
%--------------------------------------
Here, the distribution $[g(z)]_+$ is defined by
%--------------------------------------
\begin{equation}
\int_0^1 dz\, h(z)[g(z)]_+
\equiv
\int_0^1 dz \,[h(z)-h(1)]g(z).
\end{equation}
%--------------------------------------
In extracting the pole in Eq.~(\ref{D3p8}), 
we have made use of the identity,
%--------------------------------------
\begin{equation}
\label{pull-out-delta}
\frac{1}{(1-z)^{1+n\epsilon}}=
-\frac{1}{n\epsilon}\, \delta(1-z)
+
\left[
\frac{1}{(1-z)^{1+n\epsilon}}
\right]_+,
\end{equation}
%--------------------------------------
which applies when the domain of integration is $0\le z\le 1$.
The expression in Eq.~(\ref{eq:fz}) gives the exact $\epsilon$
dependence. We can expand the plus function 
$[1/(1-z)^{1+n\epsilon}]_+$ as
%--------------------------------------
\begin{equation}
\label{expand-plus}
\left[\frac{1}{(1-z)^{1+n\epsilon}}\right]_+
=
\sum_{k=0}^\infty \frac{(-n\epsilon)^k}{k!}
\left[\frac{\log^k(1-z)}{1-z}\right]_+.
\end{equation}
%--------------------------------------

In the analysis of $f(z)$, we need to keep only terms through order
$\epsilon^1$. Then, we can simplify $f(z)$ as follows:
%--------------------------------------
\begin{eqnarray}
\label{eq:fze}
f(z)\!&=&\!\!
\left(\frac{1}{1-z}\right)_+\!\!
-2\epsilon\left[\frac{\log(1-z)}{1-z}\right]_+\!\!
+\frac{1}{8}
\big[\!
-8 +21z -10z^2 + 2(13 - 7 z) \log(1-z)\big]
\nonumber\\
&+&
\frac{\epsilon}{8}
\big[
(23 - 16 z) z + 5 (8 - 11 z + 4 z^2) \log(1-z) - 
       3 (13 - 7 z) \log^2(1-z)\big]+O(\epsilon^2).
       \nonumber\\
\end{eqnarray}
%%%%%%%%%%%%%%%%%%%%%%%%%%%%%%%%%%%%%%%%%%%%%%%%%%%%%%%%%%%%%%%%%%%%%%
\subsection{$D_4[g\to Q\bar Q({}^3S_1^{[1]})]$
\label{sec:singlet-3s1}}
%%%%%%%%%%%%%%%%%%%%%%%%%%%%%%%%%%%%%%%%%%%%%%%%%%%%%%%%%%%%%%%%%%%%%%
\begin{figure}%[ht]
\begin{center}
\includegraphics[height=5cm]{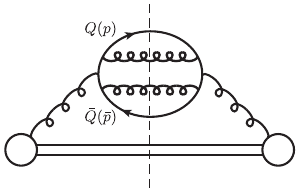}
\end{center}
\vspace{-.5cm}
\caption{
One of the 36 Feynman diagrams for the fragmentation process $g\to
Q\bar{Q}({}^3S_1^{[1]})$. Thirty-five additional diagrams can be
obtained by permuting the gluon-fermion vertices on the
left and right sides of the cut.
\label{fig:frag3}}
\end{figure}
%==============================================================
The diagrams for gluon fragmentation into a ${}^3S_1$ color-singlet 
$Q\bar Q$ pair at LO in $\alpha_s$ are shown in Fig.~\ref{fig:frag3}.
Through relative order $v^4$, the relevant squared amplitudes are
%--------------------------------------
\begin{subequations}
\label{vn-integrand}
\begin{eqnarray}
\mathcal{A}_0({}^3S_1^{[1]})&=&C_{\rm frag}\,
|{\cal M}_0({}^3S_1^{[1]})|^2,
\label{v0-integrand}
\\[1.5ex]
\mathcal{A}_2({}^3S_1^{[1]})&=&
2\,C_{\rm frag}\,
{\rm Re}[{\cal M}_{2}({}^3S_1^{[1]}){\cal M}_0^*({}^3S_1^{[1]})],
\label{v2-integrand}
\\[1.5ex]
\mathcal{A}_4({}^3S_1^{[1]})&=&
C_{\rm frag}\left\{
|{\cal M}_2({}^3S_1^{[1]})|^2+2{\rm Re}[{\cal M}_4({}^3S_1^{[1]})
{\cal M}_0^*({}^3S_1^{[1]})]\right\}.
\label{v4-integrand}
\end{eqnarray}
\end{subequations}
%--------------------------------------
The order-$v^0$ contribution $\mathcal{A}_0({}^3S_1^{[1]})$ and the 
order-$v^2$ contribution $\mathcal{A}_2({}^3S_1^{[1]})$ have been 
computed previously in Ref.~\cite{Bodwin:2003wh}.
Here, we wish to compute the order-$v^4$ 
contribution $\mathcal{A}_4({}^3S_1^{[1]})$.

The integration of $\mathcal{A}_4({}^3S_1^{[1]})$ over the phase
space contains soft divergences. These can arise when one or both of the
final-state gluons become soft. We identify the divergent part of the
integrand in Eq.~(\ref{v4-integrand}) that arises when both gluons 
become soft by making the substitutions 
$k_1\to k_1\lambda$ and $k_2\to k_2\lambda$,
multiplying by $\lambda^4$, and taking the limit $\lambda\to 0$. 
The result is 
%--------------------------------------
\begin{equation}
\mathcal{S}_{12}=
\frac{C_{{}_{\!\mathcal{S}}}(d-2)}{e_1^4 e_2^4}
\left[
(d-2)e_1^2 e_2^2
-2e_1 e_2 \,x
+x^2
\right],
\end{equation} 
%--------------------------------------
where
%--------------------------------------
\begin{equation}
C_{{}_{\!\mathcal{S}}}
=\frac{k^{+}\pi^2\alpha_s^3 z^{d-3}}{(d-1)^2(d-2)m^{8}}
\,
\frac{N_c^2-4}{4N_c^2}
\left(\frac{\mu^2}{4\pi}e^{{{\gamma}_{}}_{\rm E}}\right)^{3\epsilon}
\frac{\bm{q}^4}{m^4}.
\end{equation}
%--------------------------------------
We identify the divergent part of the integrand in
Eq.~(\ref{v4-integrand}) that arises when only $k_1$ ($k_2$) becomes soft by
subtracting $\mathcal{S}_{12}$, making the substitution $k_1\to k_1\lambda$
($k_2\to k_2\lambda$), multiplying by $\lambda^2$, and taking the limit
$\lambda\to 0$. The result is
%--------------------------------------
\begin{subequations}
\begin{eqnarray}
\mathcal{S}_{1}&=&
\frac{C_{{}_{\!\mathcal{S}}}}{e_1^4 e_2^4(1+2e_2)^2}
\sum_{n=0}^2 x^n\, h_{n}(e_1,e_2,z_1,z_2)  ,
\\
\mathcal{S}_{2}&=&
\frac{C_{{}_{\!\mathcal{S}}}}{e_2^4 e_1^4(1+2e_1)^2 }
\sum_{n=0}^2 x^n \,h_{n}(e_2,e_1,z_2,z_1) ,
\end{eqnarray}
\end{subequations}
%--------------------------------------
where the
$h_{n}(e_1,e_2,z_1,z_2)$ are given by
%--------------------------------------
\begin{subequations}
\begin{eqnarray}
h_{0}(e_1,e_2,z_1,z_2)&=&
2  e_2  (1 + 2  e_2)  \big\{
-  2  (3 - 2    \epsilon)  (1 -   \epsilon)  e_1^2  e_2^2
+z_1      e_1  e_2^2  
+z_2      e_1^2  e_2 \nonumber\\
&&\times [1 - 2  (1 - 2    \epsilon) e_2 ]
-z_1^2    e_2^2        ( 1 +     \epsilon  e_2)
+z_1  z_2  e_1  e_2      [ 1 - 2  (1 -   \epsilon)  e_2 ]
\nonumber\\
&&
+z_2^2    e_1^2        [- 1 +  (1 - 3    \epsilon)  e_2 
+ 2  (1 - 2    \epsilon)  e_2^2 \,]\,
   \big\}
+ 4  (3-2    \epsilon)  (1-  \epsilon)  e_1^2  e_2^4 
,\nonumber\\
\\
h_{1}(e_1,e_2,z_1,z_2)&=&
2  e_2  (1 + 2  e_2) 
  \big\{
z_1      e_2    (1 + 2 \epsilon  e_2    )
+ z_2      e_1    [1 + (3 - 2     \epsilon)  e_2]  
- z_1  z_2        ( 1 + 2    \epsilon e_2 
\nonumber\\
&& + 2    \epsilon e_2^2  )
- z_2^2     e_1   [3 - 2    \epsilon + 2  (1 -  \epsilon)  e_2]
\big\}
+2  e_1  e_2^2   [5 - 4    \epsilon + 6  (1 -   \epsilon)  e_2]
,\nonumber\\
\\
h_{2}(e_1,e_2,z_1,z_2)&=&
-(1 + 2 e_2) \big\{
 e_2    (4 - 3     \epsilon + 2     \epsilon   e_2)
+ 2  z_2  e_2  (1 - 2     \epsilon - 2    \epsilon    e_2)
\nonumber\\
&&-  2  z_2^2  (1 + e_2)  (1 -  \epsilon -   \epsilon  e_2)
\big\}
-(2 -   \epsilon)  e_2 
.
\end{eqnarray}
\end{subequations}
%--------------------------------------

We carry out the integrations of $\mathcal{S}_{12}$, $\mathcal{S}_1$ and
$\mathcal{S}_2$ over the phase space $d\tilde\Phi_2$ in
Eq.~(\ref{dPhi2-tilde-relation}) [see also Eq.~(\ref{dPhi2-org})] by
making use of the methods that are described in
Appendix~\ref{app:integrals}.
Then, we obtain
%--------------------------------------
\begin{eqnarray}      
\label{5-i12-ans}         
I[\mathcal{S}_{12}]
&=&
\left\{\frac{1}{8 \epsilon_{\rm IR}^2}\,\delta(1-z)
-\frac{1}{2 \epsilon_{\rm IR}}\left[\frac{1}{(1-z)^{1+4\epsilon}}\right]_{+}
+\frac{1-z^{1+2\epsilon}}{2 \epsilon_{\rm IR}(1-z)^{1+4\epsilon}}
\right\}\nonumber\\
&&\times
\left(
\frac{8\alpha_s}{3\pi m^{2}}
\right)^2
\frac{N_c^2-4}{16N_c^2}
\,
\frac{\pi\alpha_s}{(d-1) m^3}
\left(\frac{\mu^2}{4\pi}e^{{{\gamma}_{}}_{\rm E}}\right)^{\epsilon}
\frac{\bm{q}^4}{d-1}
\nonumber\\
&&\times
\left(\frac{\mu^2}{4m^2}e^{{{\gamma}_{}}_{\rm E}}\right)^{2\epsilon}
\frac{\Gamma^{2}(1+\epsilon)\Gamma^{2}(1-2\epsilon)}
{\Gamma(1-4\epsilon)}
(1-\epsilon)(6-2\epsilon-\epsilon^2-2\epsilon^3)
,
\end{eqnarray}
%--------------------------------------
where we have used the identity in
Eq.~(\ref{pull-out-delta}). 
If we expand Eq.~(\ref{5-i12-ans}) in powers of $\epsilon$, then we 
find that
%--------------------------------------
\begin{eqnarray}      
\label{5-i12-anseps}         
I[\mathcal{S}_{12}]&=&
\bigg\{
\frac{1}{8 \epsilon_{\rm IR}^2}\,\delta(1-z)
-\frac{1}{2\epsilon_{\rm IR}}
\left[
\delta(1-z)\left(
\frac{1}{3}-\log\frac{\mu}{2m}
\right)
+
\left( 
\frac{1}{1-z}
\right)_{+}
\!\!
-1
\right]
\nonumber\\
&&\;
+
\delta(1-z)\bigg(
\frac{1-3\pi^2}{48}
-\frac{2}{3}\log\frac{\mu}{2m}
+\log^2 \frac{\mu}{2m}
\bigg)
+
\left( 
\frac{1}{1-z}
\right)_{+}
\!\!
\left(
\frac{2}{3}-2\log\frac{\mu}{2m}
\right)
\nonumber\\
&&\;
+2
\left[ 
\frac{\log(1-z)}{1-z}
\right]_{+}\!\!
-2
\left(
\frac{1}{3} -\log\frac{\mu}{2m}
\right)
-\frac{z\log z}{1-z}-2\log(1-z)
\bigg\}\nonumber\\
%%%%%%%%%%%%%%%%%%
&&\times
\left(
\frac{8\alpha_s}{3\pi m^{2}}
\right)^2
\frac{N_c^2-4}{16N_c^2}
\,
\frac{\pi\alpha_s}{(d-1) m^3}
\left(\frac{\mu^2}{4\pi}e^{{{\gamma}_{}}_{\rm E}}\right)^{\epsilon}
\frac{6\bm{q}^4}{d-1}
+O(\epsilon).
\end{eqnarray}
%--------------------------------------
We also find that
%--------------------------------------
\begin{eqnarray}
I[\mathcal{S}_{1}]&=&I[\mathcal{S}_{2}]=
\left(-
\frac{\tau_1}{2\epsilon_{\rm IR}}+\tau_0
\right)
\left(\frac{\mu^2}{4m^2}e^{{{\gamma}_{}}_{\rm E}}\right)^{2\epsilon}
\frac{z^{-2+2\epsilon}(1-z)^{-4\epsilon}\Gamma^2(1+\epsilon)}
{48(1-\epsilon)}
\nonumber\\
&\times&
\left(
\frac{8\alpha_s}{3\pi m^2}
\right)^2
\frac{N_c^2-4}{16N_c^2}
\left(\frac{\mu^2}{4\pi}e^{{{\gamma}_{}}_{\rm E}}\right)^{\epsilon}
\frac{\pi\alpha_s}{(d-1)m^3}
\frac{\bm{q}^4}{d-1}+O(\epsilon),
\end{eqnarray}
%--------------------------------------
where
%--------------------------------------
\begin{subequations}
\begin{eqnarray}
\tau_0&=&
3 \bigg\{
12 (13-7 z) z^2\, \text{Li}_2(1-z)+3 (7 z-13) z^2 \log ^2(1-z)\nonumber\\
&&-2 z
\left[z^3-(43+7 \pi ^2) z^2+6 (2 z+9) z^2 \log z+(48+13 \pi
   ^2) z-42
\right]\nonumber\\
&&+3 (8 z^4-7 z^3+34 z^2-44 z+28) \log (1-z)
\bigg\}
,\\
{}\tau_1&=&
18z^2\left[
z(21-10z)+2(13-7z)\log(1-z)
\right].
\end{eqnarray}
\end{subequations}
%--------------------------------------
Here, $\text{Li}_2(z)$ is the Spence function, which is defined by
%--------------------------------------
\begin{equation}
\text{Li}_2(x)=-\int_0^x\,dt\,
\frac{\log(1-t)}{t}
=\sum_{k=1}^\infty \frac{x^k}{k^2}.
\end{equation}
%--------------------------------------
 
The fragmentation-function contribution of
$\mathcal{A}_4({}^3S_1^{[1]})$ is then
%--------------------------------------
\begin{equation}
\label{D4-tot-tilde}
\tilde{D}_4[g\to Q\bar Q({}^3S_1^{[1]})]=
\tilde{D}_4[g\to Q\bar Q({}^3S_1^{[1]})]^{\rm 
finite}+ I[\mathcal{S}_{12}]+I[\mathcal{S}_1]+I[\mathcal{S}_2],
\end{equation}
%--------------------------------------
where 
%--------------------------------------
\begin{equation}
\tilde{D}_{4}[g\to Q\bar Q({}^3S_1^{[1]})]^{\rm finite}=
\int d\tilde\Phi_2 [\mathcal{A}_4({}^3S_1^{[1]})
-\mathcal{S}_{12}-\mathcal{S}_1-\mathcal{S}_2].
\label{tilde-D4-finite}
\end{equation}
%--------------------------------------
Since $\tilde{D}_4[g\to Q\bar Q({}^3S_1^{[1]})]^{\rm finite}$ contains no soft 
divergences, we compute it in $d=4$ dimensions, using numerical 
integration over the phase space. This computation is described in 
Sec.~\ref{sec:numerical}.

As was mentioned previously, 
$D_4[g\to Q\bar Q({}^3S_1^{[1]})]^{\rm finite}$ also
contains contributions that arise from the difference between the 
fully relativistic phase space $d\Phi_2$ and the order-$v^0$ phase 
space $d\tilde\Phi_2$ [Eq.~(\ref{dPhi2-tilde-relation})]. 
Hence, we write
%--------------------------------------
\begin{equation}
\label{D4-tot}
D_4[g\to Q\bar Q({}^3S_1^{[1]})]=
D_4[g\to Q\bar Q({}^3S_1^{[1]})]^{\rm 
finite}+ I[\mathcal{S}_{12}]+I[\mathcal{S}_1]+I[\mathcal{S}_2],
\end{equation}
%--------------------------------------
where
%--------------------------------------
\begin{eqnarray}
D_4[g\to Q\bar Q({}^3S_1^{[1]})]^{\rm finite}
&=&
\tilde{D}_4[g\to Q\bar Q({}^3S_1^{[1]})]^{\rm finite}
+\frac{5\bm{q}^2}{2m^2}
\tilde{D}_2[g\to Q\bar Q({}^3S_1^{[1]})]
\nonumber\\&&
+\frac{15\bm{q}^4}{8m^4}
\tilde{D}_0[g\to Q\bar Q({}^3S_1^{[1]})],
\label{D4-finite}
\end{eqnarray}
%--------------------------------------
%--------------------------------------
\begin{subequations}
\begin{eqnarray}
\tilde{D}_0[g\to Q\bar Q({}^3S_1^{[1]})]&=&
\int \! d\tilde\Phi_2 \,  \mathcal{A}_0({}^3S_1^{[1]}),\label{tilde-D0}\\
\tilde{D}_2[g\to Q\bar Q({}^3S_1^{[1]})]&=&
\int \! d\tilde\Phi_2 \, \mathcal{A}_2({}^3S_1^{[1]}),\label{tilde-D2}
\end{eqnarray}
\end{subequations}
%--------------------------------------
and we have used the expansion
%--------------------------------------
\begin{equation}
\left(\frac{E}{m}\right)^5
=1+\frac{5\bm{q}^2}{2m^2}+\frac{15\bm{q}^4}{8m^4}+O(\bm{q}^6/m^6).
\end{equation}
%--------------------------------------
$\tilde{D}_0[g\to Q\bar Q({}^3S_1^{[1]})]$ and $\tilde{D}_2[g\to Q\bar
Q({}^3S_1^{[1]})]$ are finite, and they have been evaluated in  $d=4$
dimensions in Ref.~\cite{Bodwin:2003wh}. We have checked those
calculations, computing the phase-space integrations in
$\tilde{D}_0[g\to Q\bar Q({}^3S_1^{[1]})]$ and $\tilde{D}_2[g\to Q\bar
Q({}^3S_1^{[1]})]$ numerically.
%%%%%%%%%%%%%%%%%%%%%%%%%%%%%%%%%%%%%%%%%%%%%%%%%%%%%%%%%%%%%%%%%%%%%%
%%%%%%%%%%%%%%%%%%%%%%%%%%%%%%%%%%%%%%%%%%%%%%%%%%%%%%%%%%%%%%%%%%%%%%
\section{NRQCD LDMEs \label{sec:LDMEs}}
%%%%%%%%%%%%%%%%%%%%%%%%%%%%%%%%%%%%%%%%%%%%%%%%%%%%%%%%%%%%%%%%%%%%%%
%%%%%%%%%%%%%%%%%%%%%%%%%%%%%%%%%%%%%%%%%%%%%%%%%%%%%%%%%%%%%%%%%%%%%%
In this section we compute the NRQCD LDMEs for free $Q\bar
Q({}^3S_1^{[1]})$ states that are relevant through relative order $v^4$.
These computations are carried out in each case at the leading
nontrivial order in $\alpha_s$ in $d=4-2\epsilon$ dimensions, with 
dimensional-regularization scale $\mu$.
We remind the reader that, because we use the $\overline{\rm MS}$
scheme in computing the QCD corrections to NRQCD LDMEs, there is a
factor 
$[\mu^2\exp({{\gamma}_{}}_{\rm E})/(4\pi)]^\epsilon$ 
that is associated
with each factor of the strong coupling $\alpha_s$ in $d$ dimensions.
%%%%%%%%%%%%%%%%%%%%%%%%%%%%%%%%%%%%%%%%%%%%%%%%%%%%%%%%%%%%%%%%%%%%%%
\subsection{Order $\alpha_s^0$}
%%%%%%%%%%%%%%%%%%%%%%%%%%%%%%%%%%%%%%%%%%%%%%%%%%%%%%%%%%%%%%%%%%%%%%
The matrix elements of the $Q\bar Q$ NRQCD operators at order
$\alpha_s^0$ are normalized as 
%--------------------------------------
\begin{subequations}
\label{eq:normalization}
\begin{eqnarray}
\langle 0|
\mathcal{O}_{0}^{Q\bar{Q}({}^1S_0^{[8]})}({}^1S_0^{[8]})
|0\rangle^{(0)}
&=&(N_c^2-1),\\
\langle 0|\mathcal{O}_{0}^{Q\bar{Q}({}^3S_1^{[1]})}({}^3S_1^{[1]})
|0\rangle^{(0)}
&=&2(d-1)N_c,
\label{eq:normalization-a}
\\
\langle 0|\mathcal{O}_{0}^{Q\bar{Q}({}^3S_1^{[8]})}({}^3S_1^{[8]})
|0\rangle^{(0)}
&=&(d-1)(N_c^2-1),
\label{norm-3s18}
\\
\langle 0|
\mathcal{O}_{2}^{Q\bar{Q}({}^3S_1^{[n]})}({}^3S_1^{[n]})
|0\rangle^{(0)}
&=&\bm{q}^2
\langle 0|
\mathcal{O}_{0}^{Q\bar{Q}({}^3S_1^{[n]})}({}^3S_1^{[n]})
|0\rangle^{(0)},
\label{eq:normalization-c}
\\
\label{eq:normalization-3s1-4}
\langle 0|\mathcal{O}_{4}^{Q\bar{Q}({}^3S_1^{[n]})}({}^3S_1^{[n]})
|0\rangle^{(0)}
&=&\bm{q}^4
\langle 0|
\mathcal{O}_{0}^{Q\bar{Q}({}^3S_1^{[n]})}({}^3S_1^{[n]})
|0\rangle^{(0)},\\
\label{eq:normalization-3p8}
\langle 0|
\mathcal{O}_{0}^{Q\bar{Q}({}^3P^{[8]})}({}^3P^{[8]})
|0\rangle^{(0)}
&=& \bm{q}^2(d-1)(N_c^2-1),
\end{eqnarray}
\end{subequations}
%--------------------------------------
where a sum over the final-state polarizations is implied.
The superscript $(k)$ indicates the order in $\alpha_s$.
%%%%%%%%%%%%%%%%%%%%%%%%%%%%%%%%%%%%%%%%%%%%%%%%%%%%%%%%%%%%%%%%%%%%%%
\subsection{Order $\alpha_s$}
%%%%%%%%%%%%%%%%%%%%%%%%%%%%%%%%%%%%%%%%%%%%%%%%%%%%%%%%%%%%%%%%%%%%%%
%=====================================================================
\subsubsection{${}^3P^{[8]}\to {}^3S_1^{[1]}$}
%=====================================================================
\begin{figure}%[ht]
\begin{center}
\includegraphics[height=6cm]{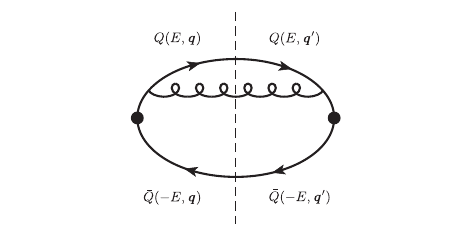}
\end{center}
\vspace{-.5cm}
\caption{
One of the four Feynman diagrams for the computation of the LDME $\langle
0|\mathcal{O}_0^{Q\bar{Q}({}^3S_1^{[1]})}
[Q\bar{Q}({}^3P^{[8]})]|0\rangle^{(1)}$. The solid circles represent the
$Q\bar Q$ operators in the LDME. As in the full-QCD calculation, we
take the free $Q$ and $\bar Q$ states to have momenta $\tfrac{1}{2}P+q$
and $\tfrac{1}{2}P-q$ on the left side of the cut and $\tfrac{1}{2}P+q'$
and $\tfrac{1}{2}P-q'$ on the right side of the cut, where,
$|\bm{q}'|=|\bm{q}|$ in the rest frame of the $Q\bar{Q}$ pair, but we
distinguish the directions of $\bm{q}$ and $\bm{q}'$ in order to be able
to project out orbital-angular-momentum states in the amplitude and its
complex conjugate. Three additional diagrams can be obtained by
permuting the gluon-fermion and operator vertices on the left and right
sides of the cut.
\label{fig:ren1}}
\end{figure}
%==============================================================
In order $\alpha_s$, the ${}^3P^{[8]}$ $Q\bar Q$ operator can couple
to the ${}^3S_1^{[1]}$ state through the diagrams that are shown in
Fig.~\ref{fig:ren1}.\footnote{We suppress Wilson lines in diagrams
involving NRQCD production operators \cite{Nayak:2005rw,Nayak:2005rt}
because the diagrams involving interactions with Wilson lines vanish for
the orders in $\alpha_s$ and the operator and final-state quantum
numbers that we consider.}

The set of diagrams in Fig.~\ref{fig:ren1} is gauge invariant. We find
it convenient to work in the $Q\bar Q$ center-of-momentum frame and to
compute the diagrams in the Coulomb gauge. Then, the real gluons in the
final state must be transverse. Through relative order $v^2$, a
straightforward computation in dimensional regularization gives
%--------------------------------------
\begin{equation}
\langle 0| {\cal O}_{0}^{Q\bar Q({}^3S_1^{[1]})}({}^3P^{[8]})
|0\rangle^{(1)}=
\left(
\mathcal{M}_a^{(1)}+\mathcal{M}_b^{(1)}+\mathcal{M}_c^{(1)}
+\mathcal{M}_d^{(1)}
\right)_{{}^3S_1^{[1]}},
\label{nrqcd-3p8}
\end{equation}
%--------------------------------------
where
%--------------------------------------
\begin{subequations}
\label{1-loop-ab}
\begin{eqnarray}
\mathcal{M}_a^{(1)}
&=&
\pi\alpha_s
\left(\frac{\mu^2}{4\pi}e^{{{\gamma}_{}}_{\rm E}}\right)^\epsilon
\frac{N_c^2-1}{4N_c^2}\,
\xi^\dagger\sigma^{k} \eta\,\eta^\dagger\sigma^{k} \xi
\nonumber\\
&&\times
\int\frac{d^{d-1}\bm{k}}{2|\bm{k}|(2\pi)^{d-1}}
\frac{
(k+2q)^l (k+2q)^i
(\delta^{ij}-\hat{\bm{k}}^i\hat{\bm{k}}^j)
(k+2q')^j (k+2q')^l 
}
{
(\bm{k}^2+2\bm{k}\cdot\bm{q}-2m|\bm{k}|)
(\bm{k}^2+2\bm{k}\cdot\bm{q}'-2m|\bm{k}|)
},\phantom{xxx}
\\
\mathcal{M}_b^{(1)}
&=&
\pi\alpha_s
\left(\frac{\mu^2}{4\pi}e^{{{\gamma}_{}}_{\rm E}}\right)^\epsilon
\frac{N_c^2-1}{4N_c^2}\,
\xi^\dagger\sigma^k \eta\,\eta^\dagger\sigma^k \xi
\nonumber\\
&&\times
\int\frac{d^{d-1}\bm{k}}{2|\bm{k}|(2\pi)^{d-1}}
\frac{
(-k+2q)^l 
(-k+2q)^i
(\delta^{ij}-\hat{\bm{k}}^i\hat{\bm{k}}^j)
(k+2q')^j
(k+2q')^l 
}
{
(\bm{k}^2-2\bm{k}\cdot\bm{q}-2m|\bm{k}|)
(\bm{k}^2+2\bm{k}\cdot\bm{q}'-2m|\bm{k}|)
},\phantom{xxx}
\\
\mathcal{M}_c^{(1)}
&=&
\pi\alpha_s
\left(\frac{\mu^2}{4\pi}e^{{{\gamma}_{}}_{\rm E}}\right)^\epsilon
\frac{N_c^2-1}{4N_c^2}\,
\xi^\dagger\sigma^k \eta\,\eta^\dagger\sigma^k \xi
\nonumber\\
&&\times
\int\frac{d^{d-1}\bm{k}}{2|\bm{k}|(2\pi)^{d-1}}
\frac{
(-k+2q)^l 
(-k+2q)^i
(\delta^{ij}-\hat{\bm{k}}^i\hat{\bm{k}}^j)
(-k+2q')^j
(-k+2q')^l 
}
{
(\bm{k}^2-2\bm{k}\cdot\bm{q}-2m|\bm{k}|)
(\bm{k}^2-2\bm{k}\cdot\bm{q}'-2m|\bm{k}|)
},\phantom{xxx}
\\
\mathcal{M}_d^{(1)}
&=&
\pi\alpha_s
\left(\frac{\mu^2}{4\pi}e^{{{\gamma}_{}}_{\rm E}}\right)^\epsilon
\frac{N_c^2-1}{4N_c^2}\,
\xi^\dagger\sigma^k \eta\,\eta^\dagger\sigma^k \xi
\nonumber\\
&&\times
\int\frac{d^{d-1}\bm{k}}{2|\bm{k}|(2\pi)^{d-1}}
\frac{
(k+2q)^l
(k+2q)^i
(\delta^{ij}-\hat{\bm{k}}^i\hat{\bm{k}}^j)
(-k+2q')^j
(-k+2q')^l
}
{
(\bm{k}^2+2\bm{k}\cdot\bm{q}-2m|\bm{k}|)
(\bm{k}^2-2\bm{k}\cdot\bm{q}'-2m|\bm{k}|)
}.\phantom{xxx}
\end{eqnarray}
\end{subequations}
%--------------------------------------
In Eq.~(\ref{1-loop-ab}), $\xi$ and $\eta^\dagger$ are the Pauli spinors
for the free $Q$ and $\bar Q$ states, respectively. The subscript
${}^3S_1^{[1]}$ in Eq.~(\ref{nrqcd-3p8}) indicates that the bispinors
$\xi^\dagger \eta$ and $\eta^\dagger \xi $ are in color-singlet,
spin-triplet states and that we project onto $S$-wave states by
averaging over the directions of $\bm{q}$ and $\bm{q}'$. A sum over
the polarizations of the spin-triplet $Q\bar{Q}$ pair is implicit.

We expand the integrands in Eq.~(\ref{1-loop-ab})
in powers of $1/m$. In dimensional regularization,
only the leading power contributes because the expressions for higher
powers in $1/m$ produce power-divergent, homogeneous
integrals.\footnote{This approach was first used in Appendix~B of
Ref.~\cite{BBL}. It has been discussed subsequently in
Refs.~\cite{Braaten:1996rp,Manohar:1997qy,Beneke:1997av}. In
Ref.~\cite{Bodwin:1998mn}, it was pointed out that this approach
allocates contributions that are infrared finite to the short-distance
coefficients, rather than to the LDMEs, and is, therefore, the NRQCD
analogue of the standard methods for computing dimensionally regulated
short-distance coefficients for hard-scattering processes in collinear
factorization in QCD.}
The result is
%--------------------------------------
\begin{eqnarray}
\langle 0| {\cal O}_{0}^{Q\bar 
Q({}^3S_1^{[1]})}({}^3P^{[8]})|0\rangle^{(1)}
&=&
\frac{8\pi \alpha_s }{m^2}
\,\frac{d-2}{d-1}
\left(\frac{\mu^2}{4\pi}e^{{{\gamma}_{}}_{\rm E}}\right)^\epsilon
\frac{N_c^2-1}{4N_c^2}
\left(\xi^\dagger q^{\prime i} q^{\prime l}\sigma^k \eta\,
\eta^\dagger q^{i} q^{l}\sigma^k 
\xi\right)_{{}^3S_1^{[1]}}
\nonumber\\
&&\times
\int\frac{d^{d-1}\bm{k}}{|\bm{k}|^3(2\pi)^{d-1}}.
\end{eqnarray}
%--------------------------------------

The remaining integration over the spatial components of $k$ is 
straightforward and yields
%--------------------------------------
\begin{eqnarray}                                                     
\langle 0| {\cal O}_{0}^{Q\bar 
Q({}^3S_1^{[1]})}({}^3P^{[8]})|0\rangle^{(1)}
=
\frac{8\alpha_sc(\epsilon)}{3\pi m^2}
\left(\frac{1}{2\epsilon_{\rm UV}}
-\frac{1}{2\epsilon_{\rm IR}}\right)
\frac{N_c^2-1}{4N_c^2}
\left(\xi^\dagger q^{\prime i} q^{\prime l}\sigma^k \eta\,
\eta^\dagger q^{i} q^{l}\sigma^k 
\xi\right)_{{}^3S_1^{[1]}}.
\label{ME3s13pj}\nonumber\\
\end{eqnarray} 
%--------------------------------------
Here, we have separated the ultraviolet (UV)
and infrared (IR) divergent
contributions of the scaleless integral. 
The quantity $c(\epsilon)$ is given by
%--------------------------------------
\begin{equation}
\label{ce}
c(\epsilon)=
\frac{(\mu^2e^{{{\gamma}_{}}_{\rm E}})^\epsilon
      (1-\epsilon)\Gamma(\tfrac{1}{2})}
      {\left(1-\frac{2}{3}\epsilon\right)(1-2\epsilon)
       \Gamma(\tfrac{1}{2}-\epsilon)}.
\end{equation}
%--------------------------------------
Note that 
%--------------------------------------
\begin{equation}
c(0)=1.
\end{equation}
%--------------------------------------
Now,
%--------------------------------------
\begin{subequations}
\begin{eqnarray}
\frac{c(\epsilon)}{\epsilon_{\rm UV}}&=&
\frac{          1}{\epsilon_{\rm UV}}
+
\frac{c(\epsilon)-1}{\epsilon},
\\
\frac{c(\epsilon)}{\epsilon_{\rm IR}}&=&
\frac{          1}{\epsilon_{\rm IR}}
+
\frac{c(\epsilon)-1}{\epsilon},
\end{eqnarray}
\end{subequations}
%--------------------------------------
where we have dropped the subscripts ``UV'' and ``IR'' in the second 
terms of the above equations because those terms are finite. 
Hence, we have
%--------------------------------------
\begin{equation}
\label{c-prod-poles}                                                    
c(\epsilon)
\left(\frac{1}{2\epsilon_{\rm UV}}
-\frac{1}{2\epsilon_{\rm IR}}\right)
=\frac{1}{2\epsilon_{\rm UV}}
-\frac{1}{2\epsilon_{\rm IR}}.
\end{equation}
%--------------------------------------
Therefore,
%--------------------------------------
\begin{eqnarray}                                                     
\langle 0| {\cal O}_{0}^{Q\bar 
Q({}^3S_1^{[1]})}({}^3P^{[8]})|0\rangle^{(1)}
=
\frac{8\alpha_s}{3\pi m^2}
\left(\frac{1}{2\epsilon_{\rm UV}}
-\frac{1}{2\epsilon_{\rm IR}}\right)
\frac{N_c^2-1}{4N_c^2}
\left(\xi^\dagger q^{\prime i} q^{\prime l}\sigma^k \eta\,
\eta^\dagger q^{i} q^{l}\sigma^k 
\xi\right)_{{}^3S_1^{[1]}}.
\label{nrqcd-3p8-int}
\nonumber\\
\end{eqnarray} 
%--------------------------------------

We renormalize $\langle 0| {\cal O}_{0}^{Q\bar
Q({}^3S_1^{[1]})}({}^3P^{[8]})|0\rangle^{(1)}$ in the 
$\overline{\rm MS}$ scheme.
In the $\overline{\rm MS}$ scheme, one constructs the counterterm for a
UV-divergent subdiagram by subtracting the poles in $\epsilon_{\rm UV}$
that appear in that subdiagram.\footnote{In some versions of the
$\overline{\rm MS}$ scheme, one subtracts constants, as well as poles,
in constructing the counterterms. These constants are accounted for in
method that we use in this paper by the factors $(\mu^2e^{\gamma_{\rm
E}})^{2\epsilon}$ that are associated with $g_s^2$.} In order to insure
that the counterterm contribution to an LDME removes precisely the
contribution that is proportional to the UV divergence in the divergent
subdiagram, one must compute all factors that are external to the
divergent subdiagram, such as the projections of external momenta onto
particular angular-momentum states, in $d=4-2\epsilon$ dimensions.
Hence, we find that the $\overline{\rm MS}$-counterterm contribution 
to $\langle 0| {\cal O}_{0}^{Q\bar Q({}^3S_1^{[1]})}({}^3P^{[8]})|0\rangle^{(1)}$ is
%--------------------------------------
\begin{eqnarray}
\delta\langle 0| {\cal O}_{0}^{Q\bar 
Q({}^3S_1^{[1]})}({}^3P^{[8]})|0\rangle^{(1)}&=&
\frac{8\alpha_s}{3\pi m^2}
\left(\frac{-1}{2\epsilon_{\rm UV}}\right)
\frac{N_c^2-1}{4N_c^2}
\left(\xi^\dagger q^{\prime i} q^{\prime l}\sigma^k \eta\,
\eta^\dagger q^i q^l\sigma^k 
\xi\right)_{{}^3S_1^{[1]}}\nonumber\\
&=&\left[Z^{(1)}({}^3P^{[8]}\to {}^3S_1^{[1]})-1\right]
\left(\xi^\dagger q^{\prime i} q^{\prime l}\sigma^k \eta\,
\eta^\dagger q^{i} q^{l}\sigma^k 
\xi\right)_{{}^3S_1^{[1]}}.\phantom{xxxxx}
\label{one-pole-counterterm}
\end{eqnarray}
%--------------------------------------
Then, we have
%--------------------------------------
\begin{eqnarray}                                                     
\langle 0| {\cal O}_{0}^{Q\bar 
Q({}^3S_1^{[1]})}({}^3P^{[8]})|0\rangle_{\overline{\rm MS}}^{(1)}
&=&
\langle 0| {\cal O}_{0}^{Q\bar
Q({}^3S_1^{[1]})}({}^3P^{[8]})|0\rangle^{(1)}
+\delta\langle 0| {\cal O}_{0}^{Q\bar
Q({}^3S_1^{[1]})}({}^3P^{[8]})|0\rangle^{(1)}
\nonumber\\
&=&
\frac{8\alpha_s}{3\pi m^2}
\left(\frac{-1}{2\epsilon_{\rm IR}}\right)
\frac{N_c^2-1}{4N_c^2}
\left(\xi^\dagger q^{\prime i} q^{\prime l}\sigma^k \eta\,
\eta^\dagger q^{i} q^{l}\sigma^k 
\xi\right)_{{}^3S_1^{[1]}}.\phantom{xxxxx}
\end{eqnarray} 
%--------------------------------------

Now we extract the $S$-wave part by averaging over the angles of 
$\bm{q}$ and $\bm{q}'$:
%--------------------------------------
\begin{eqnarray}                                                     
\langle 0| {\cal O}_{0}^{Q\bar 
Q({}^3S_1^{[1]})}({}^3P^{[8]})|0\rangle_{\overline{\rm MS}}^{(1)}
&=&
\frac{8\alpha_s}{3\pi m^2}
\left(\frac{-1}{2\epsilon_{\rm IR}}\right)
\frac{N_c^2-1}{4N_c^2}
\,
\frac{1}{d-1}
\langle 0| {\cal O}_{4,1}^{Q\bar 
Q({}^3S_1^{[1]})}({}^3S_1^{[1]})|0\rangle^{(0)}.\nonumber\\
\label{3pj3s1msbar}
\end{eqnarray} 
%=====================================================================
\subsubsection{${}^3S_1^{[8]}\to {}^3P^{[8]}$}
%=====================================================================
In order $\alpha_s$, the ${}^3S_1^{[8]}$ $Q\bar Q$ operator can couple to 
the $Q\bar Q({}^3P^{[8]})$ state. By carrying out a calculation that is very
similar to the one in the preceding section, we find that
%--------------------------------------
\begin{equation}
Z^{(1)}({}^3S_1^{[8]}\to {}^3P^{[8]})-1=
\frac{8\alpha_s}{3\pi m^2}\left(\frac{-1}{2\epsilon_{\rm UV}}\right)
\frac{N_c^2-4}{4N_c}
\label{Z3s13pj}
\end{equation}
%--------------------------------------
and that
%--------------------------------------
\begin{equation}                                                     
\langle 0| {\cal O}_{0}^{Q\bar 
Q({}^3P^{[8]})}({}^3S_1^{[8]})|0\rangle_{\overline{\rm MS}}^{(1)}
=\frac{8\alpha_s}{3\pi m^2}
\left(\frac{-1}{2\epsilon_{\rm IR}}\right)
\frac{N_c^2-4}{4N_c}
\langle 0| {\cal O}_{0}^{Q\bar 
Q({}^3P^{[8]})}({}^3P^{[8]})|0\rangle^{(0)}.\phantom{xxx}
\label{3s13pjmsbar}
\end{equation} 
%--------------------------------------
The equivalent result for NRQCD decay LDMEs was obtained 
in Ref.~\cite{Petrelli:1997ge}.
%%%%%%%%%%%%%%%%%%%%%%%%%%%%%%%%%%%%%%%%%%%%%%%%%%%%%%%%%%%%%%%%%%%%%%
\subsection{Order $\alpha_s^2$}
%%%%%%%%%%%%%%%%%%%%%%%%%%%%%%%%%%%%%%%%%%%%%%%%%%%%%%%%%%%%%%%%%%%%%%
\begin{figure}%[ht]
\begin{center}
\includegraphics[height=6cm]{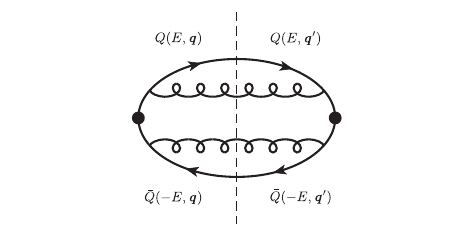}
\end{center}
\vspace{-.5cm}
\caption{
One of the 36 Feynman diagrams for computing the LDME 
$\langle0|\mathcal{O}_n^{Q\bar{Q}({}^3S_1^{[1]})}[Q\bar{Q}({}^3S_1^{[8]})]|0\rangle^{(2)}$.
 The solid circles represent
the $Q\bar Q$ operators in the LDME. Thirty-five additional diagrams can
be obtained by permuting the gluon-fermion and operator vertices on the
left and right sides of the cut.
\label{fig:ren2}}
\end{figure}
%==============================================================
In order $\alpha_s^2$, the ${}^3S_1^{[8]}$ $Q\bar Q$ operator can
couple to the $Q\bar{Q}({}^3S_1^{[1]})$ state through the diagrams
that are shown in Fig.~\ref{fig:ren2}. As in the order-$\alpha_s$ case,
we expand the integrands for these diagrams in powers of $1/m$ and,
again, only the leading power contributes in dimensional regularization.
Now we combine contributions that differ in the order of the gluon
vertices on a quark or an antiquark line by making use of the fact that
the color factor for the color-singlet contribution is symmetric under
the interchange of the gluon color indices and by making use of the
identity
%--------------------------------------
\begin{equation}
\frac{1}{|\bm{k}|}\frac{1}{|\bm{k}|+|\bm{\ell}|}
+\frac{1}{|\bm{\ell}|}\frac{1}{|\bm{k}|+|\bm{\ell}|}=
\frac{1}{|\bm{k}|}\frac{1}{|\bm{\ell}|}.
\end{equation}
%--------------------------------------
After we combine the contributions of all of the diagrams in this way, 
the loop integrations decouple, and we have
%--------------------------------------
\begin{eqnarray}
\langle 0| {\cal O}_{0}^{Q\bar 
Q({}^3S_1^{[1]})}({}^3S_1^{[8]})|0\rangle^{(2)}
&=&
\frac{(4\pi\alpha_s)^2}{m^4}
\left(
\frac{\mu^2}{4\pi}e^{{{\gamma}_{}}_{\rm E}}
\right)^{2\epsilon}
\frac{(N_c^2-4)(N_c^2-1)}{8 N_c^3}
\nonumber\\
&&\times
\left(\xi^\dagger q'^i q'^r\sigma^n \eta\,\eta^\dagger q^j q^s\sigma^n
\xi\right)_{{}^3S_1^{[1]}}
\nonumber\\
&&\times
\int\frac{d^{d-1}k}{(2\pi)^{d-1}}
\frac{
\delta^{ij}-\hat{\bm{k}}^i\hat{\bm{k}}^j
}
{
|\bm{k}|^3
}
\int\frac{d^{d-1}\ell}{(2\pi)^{d-1}}
\frac{
\delta^{rs}-\hat{\bm{\ell}}^r\hat{\bm{\ell}}^s
}
{|\bm{\ell}|^3
}\nonumber\\
&=&
\label{nrqcd-3s11-int}
\frac{1}{2}
\left[
\frac{8\alpha_s}{3\pi m^2}
\left(\frac{1}{2\epsilon_{\rm UV}}
     -\frac{1}{2\epsilon_{\rm IR}}\right)
\right]^2
\frac{(N_c^2-1)(N_c^2-4)}{16N_c^3}
\nonumber\\
&&\times
\left(\xi^\dagger q'^{i} q'^{r}\sigma^n \eta\,
\eta^\dagger q^{i} q^{r}\sigma^n
\xi\right)_{{}^3S_1^{[1]}},
\end{eqnarray}
%--------------------------------------
where we have made use of Eq.~(\ref{c-prod-poles}).

%==============================================================
\begin{figure}%[ht]
\begin{center}
\includegraphics[height=6cm]{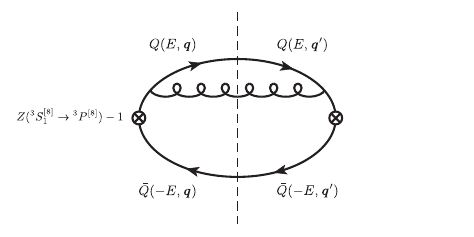}
\end{center}
\vspace{-.5cm}
\caption{One of the four Feynman diagrams for computing the one-loop
correction to the counterterm $Z({}^3S_1^{[8]} \to {}^3P^{[8]})-1$. The
symbols $\otimes$ represent the counterterm $Z({}^3S_1^{[8]}\to
{}^3P^{[8]})-1$. Three additional diagrams can be obtained by permuting
the gluon-fermion and counterterm vertices on the left and right sides
of the cut.
\label{fig:ren3}}
\end{figure}
%==============================================================
We carry out the renormalization of $\langle 0| {\cal O}_{0}^{Q\bar
Q({}^3S_1^{[1]})}({}^3S_1^{[8]})|0\rangle^{(2)}$ in the $\overline{\rm
MS}$ scheme. First, we add the contribution of the one-loop diagrams
involving the counterterm $Z^{(1)}({}^3P^{[8]}\to {}^3S_1^{[1]})-1$,
which are shown in Fig.~\ref{fig:ren3}. The contribution of these
diagrams is
%--------------------------------------
\begin{equation}
\delta_1\langle 0| 
{\cal O}_{0}^{Q\bar Q({}^3S_1^{[1]})}({}^3S_1^{[8]})
|0\rangle^{(2)}
=\left[Z^{(1)}({}^3S_1^{[8]}\to {}^3P^{[8]})-1\right]
\langle 0| {\cal O}_{0}^{Q\bar 
Q({}^3S_1^{[1]})}({}^3P^{[8]})|0\rangle^{(1)},
\end{equation}
%--------------------------------------
where $\langle 0| {\cal O}_{0}^{Q\bar                                           
Q({}^3S_1^{[1]})}({}^3P^{[8]})|0\rangle^{(1)}$ is given in 
Eq.~(\ref{nrqcd-3p8-int}) and $Z^{(1)}({}^3S_1^{[8]}\to {}^3P^{[8]})-1$ 
is given in Eq.~(\ref{Z3s13pj}). Thus,
%--------------------------------------
\begin{eqnarray}
\delta_1\langle 0| 
{\cal O}_{0}^{Q\bar Q({}^3S_1^{[1]})}({}^3S_1^{[8]})
|0\rangle^{(2)}&=&
-\frac{8\alpha_s}{3\pi m^2}
\left(\frac{1}{2\epsilon_{\rm UV}}
-\frac{1}{2\epsilon_{\rm IR}}\right)
\frac{(N_c^2-1)(N_c^2-4)}{16 N_c^3}
\nonumber\\
&&\times\frac{8\alpha_s}{3\pi m^2}\,
\frac{1}{2\epsilon_{\rm UV}}
\left(\xi^\dagger q'^i q'^r\sigma^n \eta\,\eta^\dagger q^i q^r\sigma^n
\xi\right)_{{}^3S_1^{[1]}}.
\end{eqnarray}
%--------------------------------------
Adding this counterterm contribution to
Eq.~(\ref{nrqcd-3s11-int}), we obtain
\begin{eqnarray}
&&\langle 0| 
{\cal O}_{0}^{Q\bar Q({}^3S_1^{[1]})}({}^3S_1^{[8]})
|0\rangle^{(2)}
+\delta_1\langle 0| 
{\cal O}_{0}^{Q\bar Q({}^3S_1^{[1]})}({}^3S_1^{[8]})
|0\rangle^{(2)}\nonumber\\
&&=\;
\frac{1}{8}\left(\frac{8\alpha_s}{3\pi m^2}\right)^2
\biggl(
\frac{-1}
{\epsilon_{\rm UV}^2}
+\frac{1}{\epsilon_{\rm IR}^2}\biggr)
\frac{(N_c^2-1)(N_c^2-4)}{16N_c^3}
\left(\xi^\dagger q'^{i} q'^{r}\sigma^n \eta\,
\eta^\dagger q^{i} q^{r}\sigma^n
\xi\right)_{{}^3S_1^{[1]}}.\phantom{xxx}
\label{3s11+1lcounter}
\end{eqnarray}
%--------------------------------------
We see that the counterterm contribution removes
the cross term between the pole in $\epsilon_{\rm UV}$ and the pole in
$\epsilon_{\rm IR}$. Hence, the remaining overall UV
divergence is not coupled to infrared contributions, as is 
required for the consistency of the
renormalization program. There are no single poles in $\epsilon_{\rm 
UV}$. The remaining double pole in
$\epsilon_{\rm UV}$ is removed by adding the 
$\overline{\rm MS}$-counterterm contribution
%--------------------------------------
\begin{eqnarray}
\delta_2\langle 0| 
{\cal O}_{0}^{Q\bar Q({}^3S_1^{[1]})}({}^3S_1^{[8]})
|0\rangle^{(2)}
&=&
\left(
\frac{8\alpha_s}{3\pi m^2}\right)^2
\frac{1}{8\epsilon_{\rm UV}^2}
\,
\frac{(N_c^2-1)(N_c^2-4)}{16N_c^3}
\nonumber\\
&&\times 
\left(\xi^\dagger q'^i q'^r\sigma^n \eta\,\eta^\dagger q^i q^r\sigma^n
\xi\right)_{{}^3S_1^{[1]}}\nonumber\\
&=&\left[Z^{(2)}({}^3S_1^{[8]}\to {}^3S_1^{[1]})-1\right]
\left(\xi^\dagger q'^i q'^r\sigma^n \eta\,\eta^\dagger q^i q^r\sigma^n
\xi\right)_{{}^3S_1^{[1]}}.
\nonumber\\
\end{eqnarray}
%--------------------------------------
Adding this counterterm contribution to Eq.~(\ref{3s11+1lcounter}), we obtain
%--------------------------------------
\begin{eqnarray}                                                     
\langle 0| {\cal O}_{0}^{Q\bar 
Q({}^3S_1^{[1]})}({}^3S_1^{[8]})|0\rangle_{\overline{\rm MS}}^{(2)}
&=&\langle 0| {\cal O}_{0}^{Q\bar 
Q({}^3S_1^{[1]})}({}^3S_1^{[8]})|0\rangle^{(2)}
+\delta_1\langle 0| 
{\cal O}_{0}^{Q\bar Q({}^3S_1^{[1]})}({}^3S_1^{[8]})
|0\rangle^{(2)}\nonumber\\
&&+\delta_2\langle 0| 
{\cal O}_{0}^{Q\bar Q({}^3S_1^{[1]})}({}^3S_1^{[8]})
|0\rangle^{(2)}
\nonumber\\
&=&
\left(
\frac{8\alpha_s}{3\pi m^2}\right)^2
\frac{1}{8\epsilon_{\rm IR}^2}\,
\frac{(N_c^2-1)(N_c^2-4)}{16N_c^3}
\nonumber\\
&&\times
\left(\xi^\dagger q'^i q'^r\sigma^n \eta\,\eta^\dagger q^i q^r\sigma^n
\xi\right)_{{}^3S_1^{[1]}}.
\end{eqnarray} 
%--------------------------------------
We evaluate the $S$-wave part by carrying out the average over the 
directions of $\bm{q}$ and $\bm{q}'$, with the result
%--------------------------------------
\begin{eqnarray}                                                     
\langle 0| {\cal O}_{0}^{Q\bar 
Q({}^3S_1^{[1]})}({}^3S_1^{[8]})|0\rangle_{\overline{\rm MS}}^{(2)}
&=&
\frac{1}{8\epsilon_{\rm IR}^2}
\left(
\frac{8\alpha_s}{3\pi m^2}\right)^2
\frac{(N_c^2-1)(N_c^2-4)}{16N_c^3}
\nonumber\\
&&\times
\frac{1}{d-1}\langle 0| {\cal O}_{4,1}^{Q\bar 
Q({}^3S_1^{[1]})}({}^3S_1^{[1]})|0\rangle^{(0)}.
\label{3s13s1msbar}
\end{eqnarray} 
%%%%%%%%%%%%%%%%%%%%%%%%%%%%%%%%%%%%%%%%%%%%%%%%%%%%%%%%%%%%%%%%%%%%%%
\subsection{Renormalization-group evolution}
%%%%%%%%%%%%%%%%%%%%%%%%%%%%%%%%%%%%%%%%%%%%%%%%%%%%%%%%%%%%%%%%%%%%%%
In the renormalized LDMEs, we can identify the dimensional-regularization
scale $\mu$ with the NRQCD factorization scale $\mu_\Lambda$. We now 
work out the renormalization-group evolution with respect to $\mu_\Lambda$
of the LDMEs $\langle 0| {\cal O}_{0}^{Q\bar Q({}^3S_1^{[1]})}$
$({}^3P^{[8]})|0\rangle_{\overline{\rm MS}}^{(1)}$,
$\langle 0| {\cal O}_{0}^{Q\bar Q({}^3P^{[8]})}%
({}^3S_1^{[8]})|0\rangle_{\overline{\rm MS}}^{(1)}$,
and
$\langle 0| {\cal O}_{0}^{Q\bar Q({}^3S_1^{[1]})}%
({}^3S_1^{[8]})|0\rangle_{\overline{\rm MS}}^{(2)}$.
First, we take $d/d\log{\mu_{\Lambda}}$ of Eq.~(\ref{3pj3s1msbar}): 
%--------------------------------------
\begin{eqnarray}
\frac{d}{d\log \mu_{\Lambda}}\langle 0| {\cal O}_{0}^{Q\bar Q({}^3S_1^{[1]})}
({}^3P^{[8]})|0\rangle_{\overline{\rm MS}}^{(1)}&=&
\frac{d\alpha_s}{d\log \mu_{\Lambda}}\frac{d}{d\alpha_s}
\langle 0| {\cal O}_{0}^{Q\bar Q({}^3S_1^{[1]})}         
({}^3P^{[8]})|0\rangle_{\overline{\rm MS}}^{(1)}
\nonumber\\
&=&\frac{8\alpha_s}{3\pi m^2}\,
\frac{N_c^2-1}{4N_c^2}\,
\frac{1}{d-1}
\langle 0| {\cal O}_{4,1}^{Q\bar 
Q({}^3S_1^{[1]})}({}^3S_1^{[1]})|0\rangle^{(0)},
\nonumber\\
\label{3pj3s1-evo}
\end{eqnarray}
%--------------------------------------
where we have used
%--------------------------------------
\begin{equation}
\frac{d\alpha_s}{d\log \mu_{\Lambda}}=-2\epsilon \alpha_s+O(\alpha_s^2).
\end{equation}
%--------------------------------------
The result in Eq.~(\ref{3pj3s1-evo}) agrees with the corresponding
results of Refs.~\cite{Gremm:1997dq,z-g-he}.

Similarly, taking $d/d\log{\mu_{\Lambda}}$ of Eq.~(\ref{3s13pjmsbar}), 
we obtain
%--------------------------------------
\begin{eqnarray}
\frac{d}{d\log \mu_{\Lambda}}
\langle 0| {\cal O}_{0}^{Q\bar Q({}^3P^{[8]})}
({}^3S_1^{[8]})|0\rangle_{\overline{\rm MS}}^{(1)}
&=&\frac{8\alpha_s}{3\pi m^2}
\frac{N_c^2-4}{4N_c}
\langle 0| {\cal O}_{0}^{Q\bar 
Q({}^3P^{[8]})}({}^3P^{[8]})|0\rangle^{(0)}.
\phantom{xxx}
\label{3s13pj-evo}
\end{eqnarray}
%--------------------------------------
Taking $d/d\log{\mu_{\Lambda}}$ of Eq.~(\ref{3s13s1msbar}), we also obtain
%--------------------------------------
\begin{eqnarray}                                            
\frac{d}{d\log \mu_{\Lambda}}\langle 0| {\cal O}_{0}^{Q\bar Q({}^3S_1^{[1]})}
({}^3S_{1}^{[8]})|0\rangle_{\overline{\rm MS}}^{(2)}
&=&
\left(
\frac{8\alpha_s}{3\pi m^2}\right)^2
\bigg(
\frac{-1}{2\epsilon_{\rm IR}}
\bigg)
\frac{(N_c^2-1)(N_c^2-4)}{16N_c^3}
\nonumber\\
&&\times
\frac{1}{d-1}\langle 0| {\cal O}_{4,1}^{Q\bar 
Q({}^3S_1^{[1]})}({}^3S_1^{[1]})|0\rangle^{(0)}.
\label{3s13s1-evo}
\end{eqnarray}
%--------------------------------------
Substituting Eq.~(\ref{3pj3s1msbar}) into the right side of 
Eq.~(\ref{3s13s1-evo})
we find that 
%--------------------------------------
\begin{eqnarray}
\frac{d}{d\log \mu_{\Lambda}}\langle 0| {\cal O}_{0}^{Q\bar Q({}^3S_1^{[1]})}
({}^3S_{1}^{[8]})|0\rangle_{\overline{\rm MS}}^{(2)}
&=&
\frac{8\alpha_s}{3\pi m^2}\,
\frac{N_c^2-4}{4N_c}
\langle 0| {\cal O}_{0}^{Q\bar Q({}^3S_1^{[1]})}
({}^3P^{[8]})|0\rangle_{\overline{\rm MS}}^{(1)}.
\label{3s13s1-evo2}
\end{eqnarray}
%--------------------------------------
Equations~(\ref{3s13pj-evo}) and (\ref{3s13s1-evo2}) agree with the result
in Eq.~(B19b) of Ref.~\cite{BBL} at the leading nontrivial order in $v$
and with the corresponding result in Ref~\cite{z-g-he}, but disagree
with the corresponding result in Ref.~\cite{Gremm:1997dq}.
%%%%%%%%%%%%%%%%%%%%%%%%%%%%%%%%%%%%%%%%%%%%%%%%%%%%%%%%%%%%%%%%%%%%%%
%%%%%%%%%%%%%%%%%%%%%%%%%%%%%%%%%%%%%%%%%%%%%%%%%%%%%%%%%%%%%%%%%%%%%%
\section{NRQCD matching and short-distance coefficients 
\label{sec:matching}}
%%%%%%%%%%%%%%%%%%%%%%%%%%%%%%%%%%%%%%%%%%%%%%%%%%%%%%%%%%%%%%%%%%%%%%
%%%%%%%%%%%%%%%%%%%%%%%%%%%%%%%%%%%%%%%%%%%%%%%%%%%%%%%%%%%%%%%%%%%%%%
In the discussion to follow, and in remainder of this paper, the
short-distance coefficients $d_n$ that appear are always in the
$\overline{\textrm{MS}}$ scheme. For brevity, we do not indicate the
scheme explicitly.

In relative order $v^4$, the NRQCD factorization equation that
relates full QCD and NRQCD for the fragmentation of a gluon into a 
${}^3S_1^{[1]}$ $Q\bar Q$ pair is
%--------------------------------------
\begin{eqnarray}
D_{4}[g \to Q\bar{Q}({}^3S_1^{[1]})]^{(3)}
&=&
\{
d_{4,1}[g \to Q\bar{Q}({}^3S_1^{[1]})]^{(3)}
+
d_{4,2}[g \to 
Q\bar{Q}({}^3S_1^{[1]})]^{(3)}\}
\nonumber\\&&\times\langle 0|
\mathcal{O}_{4}^{Q\bar{Q}({}^3S_1^{[1]})}({}^3S_1^{[1]})
|0\rangle^{(0)}
\nonumber\\
&+&
d_{0}[g \to Q\bar{Q}({}^3P^{[8]})]^{(2)}
\langle 0|
\mathcal{O}_{0}^{Q\bar{Q}({}^3S_1^{[1]})}({}^3P^{[8]})
|0\rangle^{(1)}
\nonumber\\
&+&
d_{0}[g \to Q\bar{Q}({}^3S_1^{[8]})]^{(1)}
\langle 0|
\mathcal{O}_{0}^{Q\bar{Q}({}^3S_1^{[1]})}({}^3S_1^{[8]})
|0\rangle^{(2)},
\label{3s11matching}
\end{eqnarray}
%--------------------------------------
where we have shown the contributions at the leading nontrivial order in 
$\alpha_s$, namely, $\alpha_s^3$, and the LDME $\langle 0|
\mathcal{O}_{4}^{Q\bar{Q}({}^3S_1^{[1]})}({}^3S_1^{[1]})
|0\rangle^{(0)}$ means either $\langle 0|
\mathcal{O}_{4,1}^{Q\bar{Q}({}^3S_1^{[1]})}({}^3S_1^{[1]})
|0\rangle^{(0)}$ or $\langle 0|
\mathcal{O}_{4,2}^{Q\bar{Q}({}^3S_1^{[1]})}$
$({}^3S_1^{[1]})
|0\rangle^{(0)}$, since they are equal at the present order
of interest in $v$, as we have explained in 
Sec.~\ref{sec:factorization}.
We wish to use this matching equation to 
determine the sum of short-distance coefficients
$d_{4,1}[g \to Q\bar{Q}({}^3S_1^{[1]})]^{(3)}
+d_{4,2}[g \to Q\bar{Q}({}^3S_1^{[1]})]^{(3)}$. 
We have already computed the $Q\bar Q$ LDMEs on the right side of
Eq.~(\ref{3s11matching}). In order to fix the short-distance
coefficients in the second and third terms on the right side of
Eq.~(\ref{3s11matching}), we make use of a matching equation at
relative order $v^2$, which is
%--------------------------------------
\begin{eqnarray}
D_2[g \to Q\bar{Q}({}^3P^{[8]})]^{(2)}
&=&
d_0[g \to Q\bar{Q}({}^3P^{[8]})]^{(2)}
\langle 0|
\mathcal{O}_{0}^{Q\bar{Q}({}^3P^{[8]})}({}^3P^{[8]})
|0\rangle^{(0)}
\nonumber\\
&+&
d_0[g \to Q\bar{Q}({}^3S_1^{[8]})]^{(1)}
\langle 0|
\mathcal{O}_{0}^{Q\bar{Q}({}^3P^{[8]})}({}^3S_1^{[8]})
|0\rangle^{(1)},
\label{3pj8matching}
\end{eqnarray}
%--------------------------------------
and a matching equation at relative order $v^0$, which is
%--------------------------------------
\begin{eqnarray}
D_0[g \to Q\bar{Q}({}^3S_1^{[8]})]^{(1)}
&=&
d_0[g \to Q\bar{Q}({}^3S_1^{[8]})]^{(1)}
\langle 0|
\mathcal{O}_{0}^{Q\bar{Q}({}^3S_1^{[8]})}({}^3S_1^{[8]})
|0\rangle^{(0)}.
\label{3s18matching}
\end{eqnarray}
%--------------------------------------
First, we solve Eq.~(\ref{3s18matching}) for $d_0[g \to 
Q\bar{Q}({}^3S_1^{[8]})]^{(1)}$, making use of Eq.~(\ref{D3s18}) for 
$D_0[g \to Q\bar{Q}({}^3S_1^{[8]})]^{(1)}$ and Eq.~(\ref{norm-3s18}) for 
$\langle 0|\mathcal{O}_{0}^{Q\bar{Q}({}^3S_1^{[8]})}({}^3S_1^{[8]})
|0\rangle^{(0)}$. The result is 
%--------------------------------------
\begin{equation}
\label{d0-3s18-1-ans}
d_0[g \to Q\bar{Q}({}^3S_1^{[8]})]^{(1)}
=
\frac{\pi\alpha_s }
{(d-1)(N_c^2-1)m^3}
\left(
\frac{\mu^2_{\Lambda}}{4\pi}e^{{{\gamma}_{}}_{\rm E}}
\right)^{\epsilon}
\delta(1-z).
\end{equation}
%--------------------------------------
The short-distance coefficient in Eq.~(\ref{d0-3s18-1-ans})
agrees with that in Refs.~\cite{BL:gfrag-NLO,Lee:2005jw}.

Next, we determine the short-distance coefficient
$d_0[g \to Q\bar{Q}({}^3P^{[8]})]^{(2)}$ by making use
of the order-$v^2$ matching equation (\ref{3pj8matching}).
We substitute 
$D_2[g \to Q\bar{Q}({}^3P^{[8]})]^{(2)}$ in Eq.~(\ref{D3p8}),
$d_0[g \to Q\bar{Q}({}^3S_1^{[8]})]^{(1)}$ 
in Eq.~(\ref{d0-3s18-1-ans}),
$\langle 0|
\mathcal{O}_{0}^{Q\bar{Q}({}^3P^{[8]})}({}^3P^{[8]})
|0\rangle^{(0)}$ in Eq.~(\ref{eq:normalization-3p8}),
and
$\langle 0|
\mathcal{O}_{0}^{Q\bar{Q}({}^3P^{[8]})}({}^3S_1^{[8]})
|0\rangle^{(1)}$ in Eq.~(\ref{3s13pjmsbar}) into
Eq.~(\ref{3pj8matching}). Then, solving for
$d_0[g \to Q\bar{Q}({}^3P^{[8]})]^{(2)}$,
we obtain
%--------------------------------------
\begin{eqnarray}
\label{d0-3p8-2-ans}
d_0[g \to Q\bar{Q}({}^3P^{[8]})]^{(2)}&=&
\frac{8\alpha_s^2}{3(d-1)(N_c^2-1)m^5}
\left(
\frac{N_c^2-4}{4N_c}
\right)
\left(
\frac{\mu^2_{\Lambda}}{4\pi}e^{{{\gamma}_{}}_{\rm E}}
\right)^{\epsilon}
\nonumber\\
&&\times
\bigg\{
-\frac{\delta(1-z)}{2\epsilon_{\rm IR}}
\left[
\frac{(1-\epsilon)\Gamma(1+\epsilon)}{1-\tfrac{2}{3}\epsilon}
\left(
\frac{\mu^2_{\Lambda}}{4m^2}e^{{{\gamma}_{}}_{\rm E}}
\right)^{\epsilon}
-1
\right]
\nonumber\\
&&
+\frac{(1-\epsilon)\Gamma(1+\epsilon)}{1-\tfrac{2}{3}\epsilon}
\left(
\frac{\mu^2_{\Lambda}}{4m^2}e^{{{\gamma}_{}}_{\rm E}}
\right)^{\epsilon}f(z)
\bigg\},
\end{eqnarray}
%--------------------------------------
where $f(z)$ is given in Eq.~(\ref{eq:fz}). 
In Ref.~\cite{Braaten:1996rp}, the color-singlet short-distance
coefficients $d_0[g \to Q\bar{Q}({}^3P_J^{[1]})]^{(2)}$ were computed.
Summing the results for $d_0[g \to Q\bar{Q}({}^3P_J^{[1]})]^{(2)}$ in
Ref.~\cite{Braaten:1996rp}
over $J=0$, 1 and 2 and multiplying by
$[(N_c^2-1)/(4N_c^2)]^{-1}[(N_c^2-4)/(4N_c)]$ in order to
obtain the corresponding short-distance coefficient for the color-octet
channel, we find agreement with our result in 
Eq.~(\ref{d0-3p8-2-ans}).

The expression in Eq.~(\ref{d0-3p8-2-ans}) gives the exact $\epsilon$
dependence. Expanding this expression to order $\epsilon^1$, using the
expression for $f(z)$ in Eq.~(\ref{eq:fze}), we find that
%--------------------------------------
\begin{eqnarray}
\label{d0-3p8-2-ans-expansion}
&&d_0[g \to Q\bar{Q}({}^3P^{[8]})]^{(2)}
\nonumber\\
&&\quad=
\frac{8\alpha_s^2}{3(d-1)(N_c^2-1)m^5}
\,
\frac{N_c^2-4}{4N_c}
\left(
\frac{\mu^2_{\Lambda}}{4\pi}e^{{{\gamma}_{}}_{\rm E}}
\right)^{\epsilon}
\nonumber\\
&&\qquad
\times
\Bigg\{
\frac{1}{6} \delta(1-z)
\bigg[1 - 6 \log\frac{\mu_{\Lambda}}{2m}
        + \epsilon\left( 
                  \frac{2}{3} - \frac{\pi^2}{4}
                + 2\log\frac{\mu_{\Lambda}}{2m} 
                -6 \log^2\frac{\mu_{\Lambda}}{2m} 
                 \right)
\bigg]
\nonumber\\
&&\qquad\phantom{xx}
    + \left(\frac{1}{1-z}\right)_+
      \left[ 1-\epsilon\left(\frac{1}{3}
      -2\log\frac{\mu_{\Lambda}}{2m} \right)
      \right]
    - 2\epsilon \left[\frac{\log(1-z)}{1-z}\right]_+
\nonumber\\
&&\qquad\phantom{xx}
    + \frac{13-7z}{4} 
      \left[ 1 - \epsilon \,\Big(\,
                     \frac{1}{3} + \frac{3}{2} \log(1-z)
                     - 2\log\frac{\mu_{\Lambda}}{2m} 
                          \,\Big) 
      \right]\log(1-z)
\nonumber\\
&&\qquad\phantom{xx}
    - \frac{1}{8}(1 - 2 z) (8 - 5 z)  
      \left(1+2\epsilon \log\frac{\mu_{\Lambda}}{2m} \right)
\nonumber\\
&&\qquad\phantom{xx}
    + \frac{\epsilon}{24}   \Big[ 8 + 48 z - 38 z^2 
              + 15 (8 - 11 z + 4 z^2) \log(1-z)
                            \Big]
\Bigg\}+O(\epsilon^2).
\end{eqnarray}
%--------------------------------------

Finally, we determine
the sum of short-distance coefficients
$d_{4,1}[g \to Q\bar{Q}({}^3S_1^{[1]})]^{(3)}
+d_{4,2}[g \to Q\bar{Q}({}^3S_1^{[1]})]^{(3)}$
by making use of the matching equation at relative order $v^4$ 
[Eq.~(\ref{3s11matching})].
We substitute 
$D_{4}[g \to Q\bar{Q}({}^3S_1^{[1]})]^{(3)}$
in Eq.~(\ref{D4-tot}),
$d_{0}[g \to Q\bar{Q}({}^3P^{[8]})]^{(2)}$
in Eq.~(\ref{d0-3p8-2-ans}),
$d_{0}[g \to Q\bar{Q}({}^3S_1^{[8]})]^{(1)}$
in Eq.~(\ref{d0-3s18-1-ans}),
$\langle 0|
\mathcal{O}_{4}^{Q\bar{Q}({}^3S_1^{[1]})}({}^3S_1^{[1]})
|0\rangle^{(0)}$
in Eq.~(\ref{eq:normalization-3s1-4}),
$\langle 0|
\mathcal{O}_{0}^{Q\bar{Q}({}^3S_1^{[1]})}({}^3P^{[8]})
|0\rangle^{(1)}$ in Eq.~(\ref{3pj3s1msbar}),
and
$\langle 0|
\mathcal{O}_{0}^{Q\bar{Q}({}^3S_1^{[1]})}({}^3S_1^{[8]})
|0\rangle^{(2)}$ in Eq.~(\ref{3s13s1msbar})
into Eq.~(\ref{3s11matching}). 
Solving for $d_{4,1}[g \to Q\bar{Q}({}^3S_1^{[1]})]^{(3)}
+d_{4,2}[g \to Q\bar{Q}({}^3S_1^{[1]})]^{(3)}$, we obtain
%--------------------------------------
\begin{eqnarray}
\label{d412sum-ans}
&&
d_{4,1}[g \to Q\bar{Q}({}^3S_1^{[1]})]^{(3)}
+d_{4,2}[g \to Q\bar{Q}({}^3S_1^{[1]})]^{(3)}
\nonumber\\
&&\qquad=\,\,
d_{4}[g \to Q\bar{Q}({}^3S_1^{[1]})]^{\rm finite}
\,\,
+\,\,\frac{2\alpha_s^3 (N_c^2 -4)}{3\pi (d - 1)^3 N_c^3  m^7 }
\,\,\bigg\{\,\,
\delta(1-z)\,
\bigg(
\frac{1}{24}
-\frac{\pi^2}{6}
\nonumber\\
&&\phantom{xxxxx}
-\frac{1}{3}\log\frac{\mu_{\Lambda}}{2m}
+\log^2\frac{\mu_{\Lambda}}{2m}
\bigg)
+\left(
\frac{1}{1-z}
\right)_{\!\!+}\!\!
\bigg(
\frac{1}{3}-2\log\frac{\mu_{\Lambda}}{2m}
\bigg)
+2\left[
\frac{\log(1-z)}{1-z}
\right]_{\!+}\!
\nonumber\\
&&\phantom{xxxxx}
-\frac{104  - 29 z - 10 z^2 }{24}
+\frac{7[z+(1+z)\log(1-z)]}{2z^2}   
   + \frac{(1-2 z) (8-5 z) }{4}\log\frac{\mu_{\Lambda}}{2m}
\nonumber\\
&&\phantom{xxxxx}
   + \frac{1 + z}{4} \left(31 - 6 z - \frac{36}{z}\right) 
        \log(1 - z)
   - \frac{z}{4}  \left(\! 39 - 6 z +  \frac{8}{1-z}
                       \!\right) \log z
\nonumber\\
&&\phantom{xxxxx}
   + \frac{13-7z}{2}\bigg[\,
              \bigg(\!
                 \log \frac{1-z}{z^2}- \log\frac{\mu_{\Lambda}}{2m}
                 \bigg) \log(1-z)   -\textrm{Li}_2(z)\,
             \bigg]\,\,   
\bigg\}+O(\epsilon),
\end{eqnarray}
%--------------------------------------
where we have used the identity 
$\textrm{Li}_2(1-z)=\pi^2/6-\textrm{Li}_2(z)-\log z\,\log(1-z)$.
$d_4[g\to Q\bar Q({}^3S_1^{[1]})]^{\rm finite}$ is defined by
%--------------------------------------
\begin{equation}                                                 
d_4[g\to Q\bar Q({}^3S_1^{[1]})]^{\rm finite}=               
\frac{D_4[g\to Q\bar Q({}^3S_1^{[1]})]^{\rm finite}}
{\langle 0|\mathcal{O}_{4}^{Q\bar{Q}({}^3S_1^{[1]})}({}^3S_1^{[1]})
|0\rangle^{(0)}}
=\frac{D_4[g\to Q\bar Q({}^3S_1^{[1]})]^{\rm finite}}
{2(d-1)N_c \bm{q}^4},
\label{d4-finite}
\end{equation}
%--------------------------------------
where $D_4[g\to Q\bar Q({}^3S_1^{[1]})]^{\rm finite}$ and
$\langle 0|\mathcal{O}_{4}^{Q\bar{Q}({}^3S_1^{[1]})}({}^3S_1^{[1]})
|0\rangle^{(0)}$ are given in Eqs.~(\ref{D4-finite}) and 
(\ref{eq:normalization-3s1-4}), respectively.
The result in Eq.~(\ref{d412sum-ans}) is new. As expected, both the 
double and single poles in $\epsilon_{\rm IR}$ have cancelled in 
Eq.~(\ref{d412sum-ans}). These cancellations rely nontrivially on  
the correctness of the infrared subtractions and on the NRQCD operator 
renormalizations in our calculation.

Finally, we note that the color-singlet short-distance coefficient
$d_0[g \to Q\bar{Q}({}^1S_0^{[1]})]^{(2)}$ was computed in
Ref.~\cite{Braaten:1996rp}. We can obtain the corresponding color-octet
short-distance coefficient by multiplying by the ratio of the 
color-octet and color-singlet color factors, namely,
$[(N_c^2-1)/(4N_c^2)]^{-1}_{N_c=3}[(N_c^2-4)/(4N_c)]$ . 
Then, we find that
%--------------------------------------
\begin{equation}
d_0[g \to Q\bar{Q}({}^1S_0^{[8]})]^{(2)}
=
\frac{\alpha_s^2}{8m^3}
\frac{N_c^2-4}{4 N_c}
\left[3z-2z^2+2(1-z)\log(1-z)\right].
\end{equation}
%%%%%%%%%%%%%%%%%%%%%%%%%%%%%%%%%%%%%%%%%%%%%%%%%%%%%%%%%%%%%%%%%%%%%%
%%%%%%%%%%%%%%%%%%%%%%%%%%%%%%%%%%%%%%%%%%%%%%%%%%%%%%%%%%%%%%%%%%%%%%
\section{Numerical results \label{sec:numerical}}
%%%%%%%%%%%%%%%%%%%%%%%%%%%%%%%%%%%%%%%%%%%%%%%%%%%%%%%%%%%%%%%%%%%%%%
%%%%%%%%%%%%%%%%%%%%%%%%%%%%%%%%%%%%%%%%%%%%%%%%%%%%%%%%%%%%%%%%%%%%%%
In this section we describe the numerical results that derive from
our calculations.

\begin{figure}%[ht]
\begin{center}
\includegraphics[width=0.5\columnwidth]{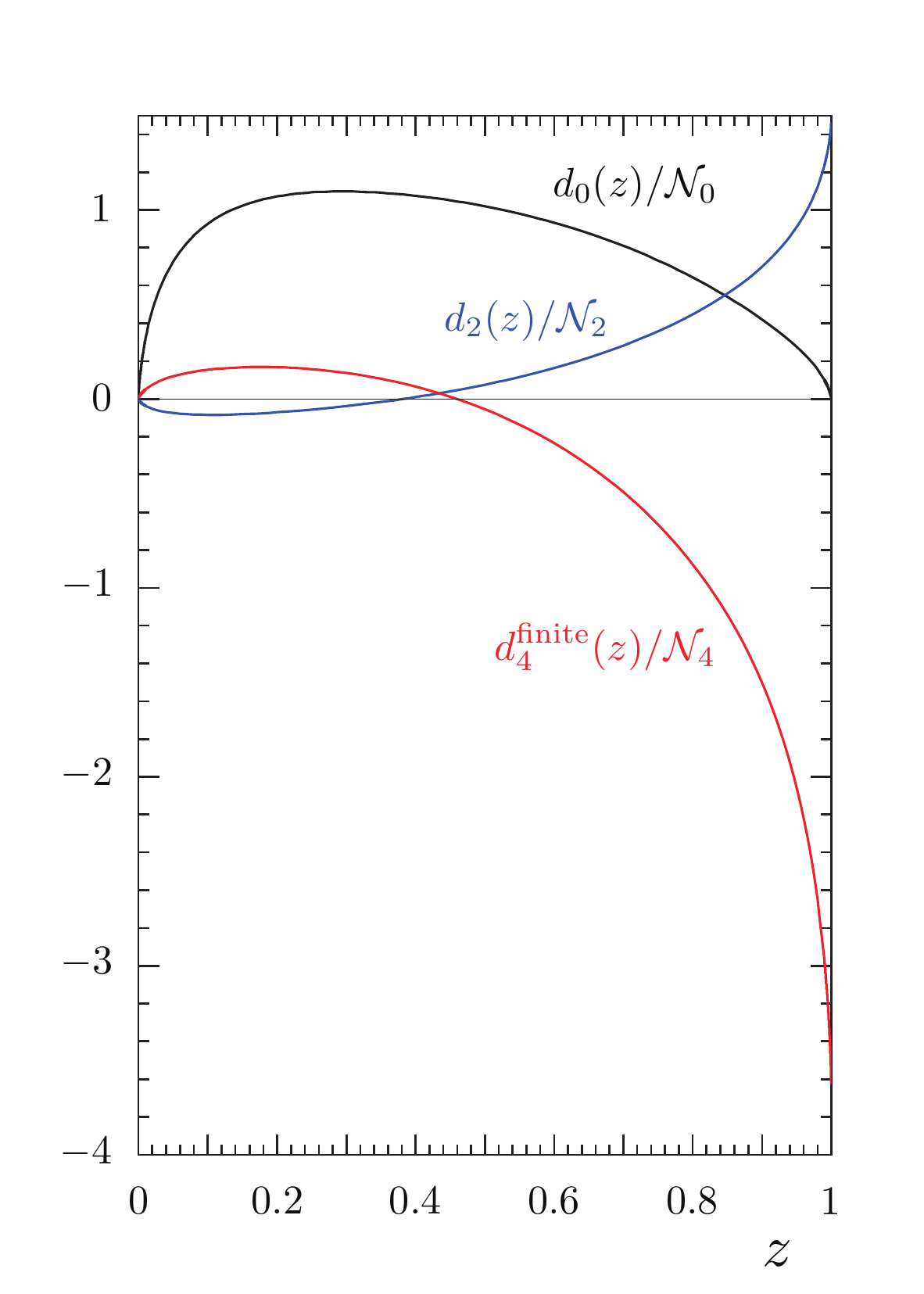}
\end{center}
\vspace{-.5cm}
\caption{The color-singlet short-distance coefficients
$d_0[g \to Q\bar{Q}({}^3S_1^{[1]})]^{(3)}(z)$,
$d_2[g \to Q\bar{Q}({}^3S_1^{[1]})]^{(3)}(z)$ and
$d_4[g \to Q\bar{Q}({}^3S_1^{[1]})]^{\rm finite}(z)$,
which are defined in Eqs.~(\ref{D0-J/psi}),
(\ref{D2-J/psi}) and (\ref{d412sum-ans}), respectively,
as functions of $z$.
The scaling factors  are 
$(\mathcal{N}_0,\mathcal{N}_2,\mathcal{N}_4)
=(10^{-3}\times\alpha_s^3/m^{3},
10^{-2}\times\alpha_s^3/m^{5},
10^{-2}\times\alpha_s^3/m^{7})$.
\label{fig8}}
\end{figure}

We have evaluated $d_{4}[g\to Q\bar Q({}^3S_1^{[1]})]^{\rm
finite}(z)$ in Eq.~(\ref{d4-finite}) by carrying out the
integrations over the phase space $d\tilde\Phi_2$ in
Eqs.~\eqref{tilde-D4-finite}, \eqref{tilde-D0}, and \eqref{tilde-D2}
numerically at $100$ points in each of the ranges $z=0$ to $z=10^{-2}$,
$z=10^{-2}$ to $z=1-10^{-2}$, and $z=1-10^{-2}$ to $z=1$.
We have then used the parametrization in
Eq.~(\ref{eq:param}) to obtain a best fit to the numerical results,
which leads to the parameters that are given in Table~\ref{tab:dz}.
Details of this procedure are given in Appendix~\ref{app:param}. We have
also applied this procedure to $d_0[g\to Q\bar
Q({}^3S_1^{[1]})]^{(3)}(z)$ and $d_2[g\to Q\bar
Q({}^3S_1^{[1]})]^{(3)}(z)$. In Fig.~\ref{fig8}, we plot the results 
of the numerical calculations of the short-distance coefficients.

%==============================================================
%==============================================================

We would like to obtain estimates of the relative sizes of the 
contributions of the various fragmentation functions to the cross 
section for $J/\psi$ production in hadron-hadron collisions. The 
contribution of a fragmentation process to the $J/\psi$ production cross 
section, differential in the $J/\psi$ transverse momentum 
${{p}_{}}_{T}$, is
%--------------------------------------
\begin{equation}
\frac{d\sigma_{J/\psi}^{\rm frag}}{d{{p}_{}}_{T}}=
\int_0^1 \! dz \, \frac{d\sigma_{g}}{d{{p}_{}}_{T}}({{p}_{}}_{T}/z)
\, D(z),
\end{equation}
%--------------------------------------
where $\frac{d\sigma_{g}}{d{{p}_{}}_{T}}({{p}_{}}_{T})$
is the cross section to produce a gluon 
with transverse momentum ${{p}_{}}_{T}$. Let us assume that 
$\frac{d\sigma_{g}}{d{{p}_{}}_{T}}({{p}_{}}_{T})\propto%
1/{{p}_{}}_{T}^\kappa$,
where $\kappa$ is a fixed power. Then, 
%--------------------------------------
\begin{equation}                                                         
\frac{d\sigma_{J/\psi}^{\rm frag}}{d{{p}_{}}_{T}}
\propto I_{\kappa}(D),
\end{equation}
%--------------------------------------
where 
%--------------------------------------
\begin{equation}
I_{\kappa}(D)=\int_0^1 \! dz \, z^{\kappa}\, D(z).
\label{eq:I-kappa}
\end{equation}
%--------------------------------------
Hence, we can obtain a rough estimate of the relative
contribution of a fragmentation process to the cross section by
computing $I_{\kappa}(D)$.\footnote{The precise calculation of a
fragmentation contribution to a cross section would require that one
compute the convolution of the two-to-two partonic cross sections that
produce a final-state gluon with the fragmentation function and with the
appropriate parton distributions. Such a calculation is beyond the scope
of the present paper.}

We can estimate $\kappa$ by taking advantage of the fact that
$J/\psi$ production in the ${}^3S_1^{[8]}$ channel is 
dominated at LO in $\alpha_s$ at large $p_{_T}$ by gluon fragmentation
into a $J/\psi$ with longitudinal-momentum fraction
$z=1$ (Ref.~\cite{Cho:1995vh}). Consequently, the $p_{_T}$ dependence in this 
channel at high $p_{_T}$ is approximately that of $d\sigma_g/dp_{_T}$. 
Furthermore, since the NLO $k$~factor for this channel 
is essentially independent of $p_{_T}$ and amounts to a correction of only 
about $14\%$ at the Tevatron \cite{Gong:2008ft}, we do not expect the 
NLO corrections to change the $p_{_T}$ dependence of this channel 
significantly. Therefore, we estimate $\kappa$ by making use 
of the result for the ${}^3S_1^{[8]}$ contribution to 
$d\sigma/dp_{_T}$ at NLO in $\alpha_s$ that appears in Fig.~1(c) of    
Ref.~\cite{Butenschoen:2010rq}. Specifically, we compute $(d/d\log 
p_{_T})\log[d\sigma/dp_{_T}]$ and find that $\kappa\approx 5.2$ at 
$p_{_T}=20$~GeV.

We have computed $I_{\kappa}(d)$ for $\kappa=0$ and $5.2$, taking
$d$ to be the order-$v^0$, order-$v^2$ and order-$v^4$ fragmentation
functions for the fragmentation process $g \to Q\bar{Q}({}^3S_1^{[1]})$.
The results are shown in
Table~\ref{tab:I-pt}.
%============================================================================
\begin{table}
\caption{
Results of computing $I_{\kappa}(d)$ for the order-$v^0$,
order-$v^2$, and
order-$v^4$ fragmentation functions for the fragmentation process $g \to
Q\bar{Q}({}^3S_1^{[1]})$. 
The first two rows are the normalization integral $I_{0}(d)$ 
for the short-distance coefficients 
$d_{0}[g\to Q\bar{Q}({}^3S_1^{[1]})]^{(3)}$ [Eq.~(\ref{D0-J/psi})],
$d_{2}[g \to Q\bar{Q}({}^3S_1^{[1]})]^{(3)}$ [Eq.~(\ref{D2-J/psi})]
and $d_{4,1}[g \to Q\bar{Q}({}^3S_1^{[1]})]^{(3)} +d_{4,2}[g \to
Q\bar{Q}({}^3S_1^{[1]})]^{(3)}$ [Eq.~\eqref{d412sum-ans}].
The third and fourth rows are $I_0(d)$ times the ratio of LDMEs
$R_n\equiv\langle 0|\mathcal{O}_n({}^3S_1^{[1]})|0\rangle/%
\langle 0|\mathcal{O}_0({}^3S_1^{[1]})|0\rangle$, as is described 
in the text. The fifth and sixth rows are the integral $I_{5.2}(d)$.
The seventh and eighth rows are $I_{5.2}(d)$ times $R_n$.
In the case of $d_{4,1}[g \to Q\bar{Q}({}^3S_1^{[1]})]^{(3)}%
+d_{4,2}[g\to Q\bar{Q}({}^3S_1^{[1]})]^{(3)}$, 
we have set $\mu_{\Lambda}=m$ and $2m$.
The quantities in the table are divided by the factor
$\mathcal{N}=10^{-4}\times\alpha_s^3/m^{3}$.
\label{tab:I-pt}}
\begin{center}
\begin{tabular}{|clclclcl|}
\hline
\mbox{\phantom{x}}&
 $I_\kappa(d_n)\,\backslash \,d_n[g\to Q\bar{Q}({}^3S_1^{[1]})]^{(3)}$
  & 
\mbox{\phantom{XXX}}& 
 $d_{0}$&
\mbox{\phantom{XXX}}&
 $d_{2}$&
\mbox{\phantom{XX}}&
 $d_{4,1}+d_{4,2}$
\mbox{\phantom{x}}
\\
\hline
&$\phantom{{}_{.2}}%
I_0(d_n)|_{\mu_\Lambda=m}$
 && 8.29    && 20.3$/m^2$   &&$-37.4\phantom{9}/m^4$  \\ 
&$\phantom{{}_{.2}}%
I_0(d_n)|_{\mu_\Lambda=2m}$
 && 8.29    && 20.3$/m^2$&&\phantom{3}$-6.54/m^4$\\ 
&$\phantom{{}_{.2}}%
I_0(d_n)|_{\mu_\Lambda=m\phantom{2}} \times R_n$
 && 8.29     &&\phantom{2}3.95&&\phantom{3}$-1.41$\\ 
&$\phantom{{}_{.2}}%
I_0(d_n)|_{\mu_\Lambda=2m}\times R_n$
&&  8.29   &&\phantom{2}3.95&&\phantom{3}$-0.247$\\ 
\hline
&$I_{5.2}(d_n)|_{\mu_\Lambda=m}$
 && 0.743    && 11.2$/m^2$    &&\phantom{-3}$35.7\phantom{9}/m^4$  \\ 
&$I_{5.2}(d_n)|_{\mu_\Lambda=2m}$
 && 0.743  && 11.2$/m^2$   &&\phantom{-3}$84.5\phantom{9}/m^4$   \\ 
&$I_{5.2}(d_n)|_{\mu_\Lambda=m\phantom{2}}\times R_n$
 &&0.743  &&\phantom{1}2.18 &&\phantom{-39}1.35 \\ 
&$I_{5.2}(d_n)|_{\mu_\Lambda=2m}\times R_n$
 && 0.743   &&\phantom{1}2.18   &&\phantom{-39}3.19 \\ 
\hline
\end{tabular}
\end{center}
\end{table}
%============================================================================
In the first and second rows of Table~\ref{tab:I-pt}, we give the
normalization integrals $I_{0}(d)$ for the short-distance coefficients.
In the third and fourth rows, we give $I_{0}(d)$ multiplied by the
relative value of corresponding LDME. In the fifth and sixth rows we
give $I_{5.2}(d)$. In the seventh and eighth rows we give $I_{5.2}(d)$
multiplied by the relative value of corresponding LDME. We take the
relative values of the LDMEs to be given by the generalized
Gremm-Kapustin relation \cite{Bodwin:2006dn}:
%--------------------------------------
\begin{equation}
\langle 0|
\mathcal{O}_{0}^{J/\psi}({}^3S_1^{[1]})
|0\rangle:
\langle 0|
\mathcal{O}_{2}^{J/\psi}({}^3S_1^{[1]})
|0\rangle:
\langle 0|
\mathcal{O}_{4}^{J/\psi}({}^3S_1^{[1]})
|0\rangle
=1:
m^2\langle v^2 \rangle:
m^4\langle v^2 \rangle^2.
\end{equation}
We use the value of $m^2\langle v^2 \rangle$ from
Table I of Ref.~\cite{Bodwin:2007fz}:
\begin{equation}
m^2\langle v^2 \rangle=0.437\,\textrm{GeV}^2\,\,\,
\textrm{at}\,\,\,m=1.5\,\textrm{GeV}.
\label{v-sq}
\end{equation}
%--------------------------------------

Examining the relative values in the fifth or sixth rows of
Table~\ref{tab:I-pt}, we see that, if we exclude the factors from the
NRQCD LDMEs, then the order-$v^4$ contribution is enhanced considerably
relative to the order-$v^0$ and order-$v^{2}$ contributions. We attribute
this enhancement to the strong peaking of $d_{4,1}[g \to
Q\bar{Q}({}^3S_1^{[1]})]^{(3)} +d_{4,2}[g \to
Q\bar{Q}({}^3S_1^{[1]})]^{(3)}$ near $z=1$, which arises primarily from
the $\delta$ functions and $+$ distributions that it contains.
We expect such strong peaking to occur whenever the full-QCD process
contains soft divergences that correspond to divergences in the $Q\bar
Q$ LDMEs.

From the seventh and eighth rows of Table~\ref{tab:I-pt}, we see
that the order-$v^4$ fragmentation contribution to the cross section is
enhanced by approximately a factor of 2 for
$\mu_{\Lambda}=m$ and by approximately a factor of 4 for $\mu_{\Lambda}=2m$,
relative to the order-$v^0$ contribution. However, the order-$v^0$
contribution to the $J/\psi$ cross section at ${{p}_{}}_{T}=10$~GeV lies
about a factor of 30 below the measured cross section. Thus, while the
enhancement of the order-$v^4$ fragmentation contribution to the cross
section is substantial, it is not sufficient to make the order-$v^4$
fragmentation contribution to the cross section important for the
phenomenology of $J/\psi$ production at the current level of precision.

We can also compare the relative contributions to 
$d\sigma_{J/\psi}^{\rm frag}$ of each of the three $Q\bar Q$ 
channels in that appear in
$D[g\to J/\psi]$ in order $v^4$ [Eq.~(\ref{D4-Jpsi})]. 
For each channel, we compute $I_\kappa(5.2)$.
Then, we multiply by the following LDMEs:
%--------------------------------------
\begin{subequations}
\label{bk-LDMEs}
\begin{eqnarray}
\langle 0|\mathcal{O}_{4}^{J/\psi}({}^3S_1^{[1]})
|0\rangle&=&m^4\langle v^2\rangle^2 \times 1.32~{\rm GeV}^3,
\\
\langle 0|
\mathcal{O}_{0}^{J/\psi}({}^3P^{[8]})
|0\rangle&=&-0.109~{\rm GeV}^5,
\\
\langle 0|
\mathcal{O}_{0}^{J/\psi}({}^3S_1^{[8]})
|0\rangle&=&3.12\times 10^{-3}~{\rm GeV}^3
,
\\
\langle 0|
\mathcal{O}_{0}^{J/\psi}({}^1S_0^{[8]})
|0\rangle&=&4.50\times10^{-2}~{\rm GeV}^3
,
\end{eqnarray}
\end{subequations}
%--------------------------------------
where $m^2\langle v^2\rangle$ is given in Eq.~(\ref{v-sq}), and the 
numerical values on the right sides of Eq.~(\ref{bk-LDMEs}) come from 
the fit at NLO in $\alpha_s$ to the Tevatron and HERA data in 
Ref.~\cite{Butenschoen:2010rq}. The results of this computation are 
shown in Table~\ref{tab:v4-frag}.
%============================================================================
\begin{table}
\caption{
Relative contributions to $d\sigma_{J/\psi}^{\rm frag}$ in order $v^4$.
The first and second rows give $I_{0}(d)$ for the short-distance
coefficients in Eq.~(\ref{D4-Jpsi}). The second and third rows give
$I_{0}(d)$ times the LDMEs in Eq.~(\ref{bk-LDMEs}). The fourth and
fifth  rows give $I_{5.2}(d)$. The fifth and sixth rows give 
$I_{5.2}(d)$ times the LDMEs in Eq.~(\ref{bk-LDMEs}). We take 
$m=m_c=1.5\,\textrm{GeV}$. For compatibility with 
Ref.~\cite{Butenschoen:2010rq}, we take $\alpha_s=\alpha_s({{m}_{}}_{T})$,
where
${{m}_{}}_{T}=\sqrt{{{p}_{}}_{T}^2+4m_c^2}$. We choose the point
${{p}_{}}_{T}=20$~GeV, which implies that $\alpha_s({{m}_{}}_{T})=0.154$.
\label{tab:v4-frag}}
\begin{center}
\begin{tabular}{|llclll|}
\hline
$I_\kappa(d)\,\backslash \,$channel&
\!\!\!\!$\mathcal{O}_{0}^{J/\psi}({}^1S_0^{[8]})$&~&
\!\!\!\!$\mathcal{O}_{0}^{J/\psi}({}^3S_1^{[8]})$&
$\mathcal{O}_{0}^{J/\psi}({}^3P^{[8]})$&
$\mathcal{O}_{4}^{J/\psi}({}^3S_1^{[1]})$
\\
\hline
$\phantom{{}_{.2}}%
I_0(d)|_{\mu_\Lambda=m\phantom{2}}\times\,10^6\,\textrm{GeV}^{3}$&
$122.$&&
$5970$&
$-171.\,\textrm{GeV}^{-2}$&
$-0.799\,\textrm{GeV}^{-4}$
\\ 
$\phantom{{}_{.2}}%
I_0(d)|_{\mu_\Lambda=2m}\times\,10^6\,\textrm{GeV}^{3}$&
$122.$&&
$5970$&
$-271.\,\textrm{GeV}^{-2}$&
$-0.140\,\textrm{GeV}^{-4}$
\\ 
$\phantom{{}_{.2}}%
I_0(d)|_{\mu_\Lambda=m\phantom{2}}\times\,10^6\,\textrm{LDME}$\phantom{x}&
$\phantom{12}5.49$&&
$\phantom{59}18.6$&
$\phantom{-1}18.6$&
$-0.201$
\\
$\phantom{{}_{.2}}%
I_0(d)|_{\mu_\Lambda=2m}\times\,10^6\,\textrm{LDME}$&
$\phantom{12}5.49$&&
$\phantom{59}18.6$&
$\phantom{-1}29.6$&
$-0.0353$
\\
\hline
$I_{5.2}(d)|_{\mu_\Lambda=m\phantom{2}}\times\,10^6\,\textrm{GeV}^{3}$&
$\phantom{1}36.7$&&
$5970$&
$-300.\,\textrm{GeV}^{-2}$&
$\phantom{-}0.763\,\textrm{GeV}^{-4}$
\\
$I_{5.2}(d)|_{\mu_\Lambda=2m}\times\,10^6\,\textrm{GeV}^{3}$&
$\phantom{1}36.7$&&
$5970$&
$-400.\,\textrm{GeV}^{-2}$&
$\phantom{-}1.81\phantom{2}\,\textrm{GeV}^{-4}$
\\
$I_{5.2}(d)|_{\mu_\Lambda=m\phantom{2}}\times\,10^6\,\textrm{LDME}$&
$\phantom{12}1.65$&&
$\phantom{59}18.6$&
$\phantom{-1}32.7$&
$\phantom{-}0.192$
\\
$I_{5.2}(d)|_{\mu_\Lambda=2m}\times\,10^6\,\textrm{LDME}$&
$\phantom{12}1.65$&&
$\phantom{59}18.6$&
$\phantom{-1}43.6$&
$\phantom{-}0.455$
\\
\hline
\end{tabular}
\end{center}
\end{table}
%============================================================================

We see from the second row of Table~\ref{tab:v4-frag} that the $Q\bar
Q({}^3S_1^{[1]})$ channel makes a small contribution to $D_4[g\to
J/\psi]$ at ${p_{}}_T=20$~GeV, confirming our previous conclusion that this
channel is not important phenomenologically at the current level of
precision. We also see that the $Q\bar Q({}^3S_1^{[8]})$ and $Q\bar
Q({}^3P^{[8]})$ channels give comparable contributions to $D_4[g\to
J/\psi]$ at ${p_{}}_T=20$~GeV. As we have mentioned, at high $p_{_T}$ 
at LO in $\alpha_s$, the
fragmentation contribution gives the bulk of the contribution in the
$Q\bar Q({}^3S_1^{[8]})$ channel (Ref.~\cite{Cho:1995vh}). As we have 
also mentioned, the correction to the production rate in the $Q\bar
Q({}^3S_1^{[8]})$ channel at NLO in $\alpha_s$ is only about $14\%$ at
the Tevatron \cite{Gong:2008ft}. Hence, we can estimate the fragmentation
contribution to $J/\psi$ production at the Tevatron in the $Q\bar
Q({}^3P^{[1]})$ channel by multiplying the ratio of the $Q\bar
Q({}^3P^{[8]})$ and $Q\bar Q({}^3S_1^{[8]})$ entries in the seventh row
of Table~\ref{tab:v4-frag} by the value of $d\sigma/d{p_{}}_T\times
B(J/\psi\to \mu\mu)$ at $20$~GeV from Fig.~1(c) of
Ref.~\cite{Butenschoen:2010rq} and by dividing by $1.14$ to account
for the NLO correction in the $Q\bar Q({}^3S_1^{[8]})$ channel. Our
estimate is that the fragmentation contribution to
$d\sigma/d{p_{}}_T\times B(J/\psi\to \mu\mu)$ at ${p_{}}_T=20$~GeV from
the $Q\bar Q({}^3P^{[8]})$ channel is about $6\times
10^{-3}$~nb/GeV. We see that this is comparable to (about a factor
of $2$ larger than) the total NLO contribution in the $Q\bar
Q({}^3P^{[8]})$ channel in Fig.~1(c) of Ref.~\cite{Butenschoen:2010rq}.
A more precise calculation of the fragmentation contribution in the $Q\bar
Q({}^3P^{[8]})$ channel will be necessary in order to determine whether
it is the dominant contribution in that channel at NLO in $\alpha_s$ at
high $p_{_T}$.
%%%%%%%%%%%%%%%%%%%%%%%%%%%%%%%%%%%%%%%%%%%%%%%%%%%%%%%%%%%%%%%%%%%%%%
%%%%%%%%%%%%%%%%%%%%%%%%%%%%%%%%%%%%%%%%%%%%%%%%%%%%%%%%%%%%%%%%%%%%%%
\section{Summary\label{sec:discussion}}
%%%%%%%%%%%%%%%%%%%%%%%%%%%%%%%%%%%%%%%%%%%%%%%%%%%%%%%%%%%%%%%%%%%%%%
%%%%%%%%%%%%%%%%%%%%%%%%%%%%%%%%%%%%%%%%%%%%%%%%%%%%%%%%%%%%%%%%%%%%%%
We have calculated NRQCD short-distance coefficients for gluon
fragmentation into a ${}^3S_1$ heavy-quarkonium state through relative
order $v^4$. Our principal new result is the expression for the sum of
the relative-order-$v^4$ short-distance coefficients $d_{4,1}[g \to
Q\bar{Q}({}^3S_1^{[1]})]^{(3)} +d_{4,2}[g \to
Q\bar{Q}({}^3S_1^{[1]})]^{(3)}$ for gluon fragmentation through the
${}^3S_1$ color-singlet channel. This expression is given in
Eq.~(\ref{d412sum-ans}) and in the parametrization of $d_4[g\to
Q\bar{Q}({}^3S_1^{[1]})]^{\rm finite}$ in Eq.~(\ref{eq:param}). As a
byproduct of this calculation, we have computed the short-distance
coefficient for gluon fragmentation through the ${}^3S_1$ color-octet
channel [Eq.~(\ref{d0-3s18-1-ans})], finding agreement with the results
in Refs.~\cite{BL:gfrag-NLO,Lee:2005jw}, and the sum of short-distance
coefficients for gluon fragmentation through the ${}^3P_J$ color-octet
channel [Eq.~(\ref{d0-3p8-2-ans})], finding agreement with the result in
Ref.~\cite{Braaten:1996rp} for the ${}^3P_J$ color-singlet
short-distance coefficients. We have also computed the short-distance
coefficients for gluon fragmentation through the ${}^3S_1$ color-singlet
channel at leading order in $v$ and at relative-order $v^2$ and find
agreement with the results in Refs.~\cite{Braaten:1993rw,Braaten:1995cj}
and Ref.~\cite{Bodwin:2003wh}, respectively.

The analysis in this paper involves, for the first time in an NRQCD
factorization calculation, both single and double soft divergences in
the full-QCD process. These soft divergences manifest themselves as
single and double poles in $\epsilon=(4-d)/2$ in dimensional
regularization. We have dealt with the soft divergences by devising
subtractions that remove both the single and double poles in $\epsilon$.
We have calculated the phase-space integrals for the subtraction
contributions analytically in $d=4-2\epsilon$ dimensions and have
calculated the phase-space integrals for the finite remainder
contributions numerically in $d=4$ dimensions. We have also
calculated the perturbative NRQCD LDMEs that appear in the NRQCD
matching equations for the short-distance coefficients. These
perturbative LDMEs involve both one-loop and two-loop renormalizations
of the NRQCD operators. Our results for the renormalization-group
evolution of the renormalized LDMEs confirm some results in
Refs.~\cite{BBL,Gremm:1997dq,z-g-he}, but disagree with one of the
results in Ref.~\cite{Gremm:1997dq}.

In Tables~\ref{tab:I-pt} and \ref{tab:v4-frag},  we have given estimates
of the relative sizes of the contributions of the various channels to
gluon fragmentation into $J/\psi$ through relative order $v^4$. As we
have mentioned, the contribution of the ${}^3S_1$ color-octet channel is
believed to be important phenomenologically at large $p_{_T}$. Hence, one
might expect the order-$v^4$ contribution in the ${}^3S_1$ color-singlet
channel to be important as well, since the color-singlet channel
mixes with the color-octet channel through single and double
logarithms of the NRQCD factorization scale at order $v^4$. Indeed, we
find that the contribution to the cross section at $p_{_T}=20$~GeV of
the ${}^3S_1$ color-singlet channel in order $v^4$ is about a factor
of $2$ larger than the contribution of the ${}^3S_1$ color-singlet
channel at the leading order in $v$ when the factorization scale is taken to
be $\mu_\Lambda=m_c$ and is about a factor of $4$ larger than the
contribution of the ${}^3S_1$ color-singlet channel at the leading
order in $v$ when the factorization scale is taken to be
$\mu_\Lambda=2m_c$. This is a large enhancement, since one would
nominally expect the order-$v^4$ contribution to be about $\langle
v^2\rangle^2\approx 0.04$ times the order-$v^0$ contribution. In spite
of this large enhancement of the fragmentation contribution in order
$v^4$, the corresponding contribution to the $J/\psi$ production cross
section at the Tevatron or the LHC is not important at the current level
of precision of the phenomenology.

We attribute the large enhancement of the order-$v^4$ contribution to
the ${}^3S_1$ color-singlet channel to the peaking of the sum of
short-distance coefficients near $z=1$. This peaking arises from terms
that are proportional to $\delta(1-z)$, $[1/(1-z)]_+$, and
$[\log(1-z)/(1-z)]_+$. These terms are remnants of the soft divergences
that appear in the full-QCD expression for the fragmentation process and
that, ultimately, cancel in the NRQCD matching
equations for the short-distance coefficients. We expect such a peaking,
and the corresponding enhancement, to be present whenever soft
divergences appear in a full-QCD process and are cancelled in the NRQCD 
matching equations.\footnote{There is also some peaking/enhancement in the
order-$v^2$ contribution to the ${}^3S_1$ color-singlet channel. The
order-$v^2$ contribution contains terms that are the product of an
order-$v^0$ amplitude with an order-$v^2$ amplitude. The order-$v^2$
amplitude is sufficiently singular that its square would produce a soft
divergence. However, in combination with the order-$v^0$ amplitude, it
produces only a peaking at $z=1$.}

The signature of this peaking/enhancement in Table~\ref{tab:v4-frag} is
that the magnitude of the $I_{5.2}$ entry is comparable to or
greater than the magnitude of the $I_0$ entry. We see this signature of
peaking/enhancement in the ${}^3S_1$ and ${}^3P_J$ color-octet channels,
as well as in the ${}^3S_1$ color-singlet channel in order $v^4$, but not
in the ${}^1S_0$ color-octet channel. The ${}^3P_J$ color-octet channel
contains soft divergences in the full-QCD process, but the ${}^1S_0$
color-octet channel does not. (The fragmentation contribution in the
${}^3S_1$ color-octet channel is proportional to $\delta(1-z)$ at
leading order in $\alpha_s$.)

The fragmentation contribution to the ${}^3S_1$ color-octet channel
dominates the contribution of that channel to the $J/\psi$ production
cross section at hadron-hadron colliders at large $p_{_T}$, both at LO in
$\alpha_s$ (Ref.~\cite{Cho:1995vh}) and at NLO in $\alpha_s$
(Ref.~\cite{Gong:2008ft}). We estimate that the fragmentation
contribution to the ${}^3P_J$ color-octet channel gives a substantial
part of the total contribution of that channel at NLO in $\alpha_s$ to the
$J/\psi$ production cross section at hadron-hadron colliders at
large $p_{_T}$. A more precise calculation of the ${}^3P_J$
color-octet fragmentation contribution will be required in order to
determine if it is the dominant contribution at large $p_{_T}$ in that
channel.
%%%%%%%%%%%%%%%%%%%%%%%%%%%%%%%%%%%%%%%%%%%%%%%%%%%%%%%%%%%%%%%%%%%%%%
%%%%%%%%%%%%%%%%%%%%%%%%%%%%%%%%%%%%%%%%%%%%%%%%%%%%%%%%%%%%%%%%%%%%%%
\begin{acknowledgments}
We thank Zhi-Guo He for providing us with his unpublished results
for the renormalization-group evolution of NRQCD LDMEs.
We also thank Andrea Petrelli for the use of some of his 
\textsc{mathematica} code.
We thank An-Ping Chen
for bringing to our attention sign errors 
in Eq.~(6.3) in a previous version of this paper.
The work of G.T.B.\ in the High Energy Physics Division at Argonne
National Laboratory was supported by the U.~S.~Department of Energy,
Division of High Energy Physics, under Contract No.\
DE-AC02-06CH11357. The work of U.R.K.\ and J.L.\ was supported by
the MEST of Korea under the NRF Grants No.~2011-0027559 and
No.~2011-0003023, respectively. 
\end{acknowledgments}
% Create the reference section using BibTeX:
\appendix
%%%%%%%%%%%%%%%%%%%%%%%%%%%%%%%%%%%%%%%%%%%%%%%%%%%%%%%%%%%%%%%%%%%%%%
%%%%%%%%%%%%%%%%%%%%%%%%%%%%%%%%%%%%%%%%%%%%%%%%%%%%%%%%%%%%%%%%%%%%%%
\section{Analytic calculation of phase-space integrals\label{app:integrals}}
%%%%%%%%%%%%%%%%%%%%%%%%%%%%%%%%%%%%%%%%%%%%%%%%%%%%%%%%%%%%%%%%%%%%%%
%%%%%%%%%%%%%%%%%%%%%%%%%%%%%%%%%%%%%%%%%%%%%%%%%%%%%%%%%%%%%%%%%%%%%%
In this appendix, we give some of the details of the analytic
integrations over the final-state phase space for $S_{12}$, $S_{1}$ and
$S_{2}$.

We first carry out the average over the angles of the transverse
components of the final-state gluon momenta in the phase-space
(\ref{dPhi2-org}). Under this angular averaging we have
%--------------------------------------
\begin{subequations}
\begin{eqnarray}
\langle x\rangle_{\perp}&=& 
\frac{z_1 z_2}{z}\left(\frac{e_1}{z_1}+\frac{e_2}{z_2}-\frac{1}{z}
\right),\\
\langle x^2\rangle_{\perp}&=& 
\frac{z_1^2z_2^2}{z^2}
\left(\frac{e_1}{z_1}+\frac{e_2}{z_2}-\frac{1}{z}\right)^2
+\frac{z_1 z_2}{(d-2)z^2}\left(2e_1-\frac{z_1}{z}\right)
\left(2e_2-\frac{z_2}{z}\right).
\end{eqnarray}
\end{subequations}
%--------------------------------------
After this angular averaging, the squared amplitude is independent of the
directions of $\bar{\bm{k}}_{1\perp}$ and $\bar{\bm{k}}_{2\perp}$.

In evaluating the integrals over 
$|\bar{\bm{k}}_{1\perp}|$ and $|\bar{\bm{k}}_{2\perp}|$,
it is convenient to define
dimensionless variables $u_1$ and $u_2$:
%--------------------------------------
\begin{subequations}
\label{u1-u2}
\begin{eqnarray}
u_1&=&\left(\frac{z}{z_1}\right)^2\bar{k}^2_{1\perp},\\
u_2&=&\left(\frac{z}{z_2}\right)^2\bar{k}^2_{2\perp}.
\end{eqnarray}
\end{subequations}
%--------------------------------------
Then $e_1$, $e_2$ and $x$ can be expressed as
%--------------------------------------
\begin{subequations}
\begin{eqnarray}
e_1&=&
\frac{z_1}{2z}(1+u_1),
\\
e_2&=&
\frac{z_2}{2z}(1+u_2).
\end{eqnarray}
\end{subequations}
%--------------------------------------
The integrals of $S_{12}$, $S_{1}$ and $S_{2}$ over the magnitudes
of the transverse components of the final-state gluon momenta can be
expressed as linear combinations of the following elementary integrals:
%--------------------------------------
\begin{eqnarray}
J_{mn}^i
&=&
\int\frac{d^{d-2}\bm{\bar{k}}_{i\perp}}{(2\pi)^{d-2}}
\frac{1}{e_i^m(1+2e_i)^n}\nonumber\\
&=&
\frac{2^m}{(4\pi)^{1-\epsilon}\Gamma(1-\epsilon)}
\left(\frac{z_i}{z}\right)^{2-m-n-2\epsilon}
\int_0^\infty \frac{du_i}
{u_i^\epsilon(1+u_i)^m\left(1+\frac{z}{z_i}+ u_i\right)^n},
\phantom{xxxx}
\end{eqnarray}
where $m$ takes on integer values equal to or 
greater than $-1$, and $n$ takes on non-negative integer values.
These integrals can be evaluated straightforwardly to obtain
%--------------------------------------
\begin{subequations}
\begin{eqnarray}
\label{in-fin}
J^i_{m0}&=&
2^m
\left(\frac{z_i}{z}\right)^{2-m-2\epsilon}
\frac{\Gamma(m-1+\epsilon)}{(4\pi)^{1-\epsilon}\Gamma(m)}
,\\
\label{j0n-fin}
%----------------
J^i_{0n}&=&
\left(\frac{z_i}{z}\right)^{1-\epsilon}
\left(1+\frac{z_i}{z}\right)^{1-n-\epsilon}
\frac{ \Gamma(n-1+\epsilon)}{(4\pi)^{1-\epsilon}\Gamma(n)}
,\\
\label{j-2-fin}
J^i_{-1,2}
&=&
\frac{1}{2}
\left(
J^i_{01}
-
J^i_{02}
\right)
,\\
%----------------
\label{jmn-fin}
J_{mn}^i
&=&
(-1)^{m+n}
\left(\frac{z_i}{z}\right)^{1-\epsilon}
\frac{\Gamma(\epsilon)}{(4\pi)^{1-\epsilon}\Gamma(m)\Gamma(n)}
\nonumber\\
&&\times
\bigg(\frac{\partial}{\partial a}\bigg)^{m-1}
\bigg(\frac{\partial}{\partial b}\bigg)^{n-1}
\left\{\frac{2}{b-2a}\bigg[
\left(2a+\frac{z_i}{z}\right)^{-\epsilon}
-
\left(b+\frac{z_i}{z}\right)^{-\epsilon}
\bigg]
\right\}
\Bigg|_{a=0,\,b=1}.\phantom{xxxx}
\end{eqnarray}
\end{subequations}
%--------------------------------------
After we have carried out the integrations over the transverse 
components of the final-state gluon momenta, the remaining integrals 
over $z_1$ and $z_2$ for $S_{12}$ are simple. The integrations over 
$z_1$ and $z_2$ for $S_1$ and $S_2$ can be expressed as 
linear combinations of the integrals
%--------------------------------------
\begin{subequations}
\begin{eqnarray}
A_n&=&\int_{z_1z_2}
\frac{{z_2}^{n-\epsilon}}{{z_1}^{1+2\epsilon}(z+z_2)^{1+\epsilon}},\\
B_n&=&\int_{z_1z_2}
\frac{{z_2}^{n-\epsilon}}
{{z_1}^{1+2\epsilon}(z+z_2)^{\epsilon}},\\
C_n&=&\int_{z_1z_2}
\frac{{z_2}^{n-2\epsilon}}{{z_1}^{1+2\epsilon}},\\
D_n&=&\int_{z_1z_2}
\frac{{z_2}^{n-2\epsilon}}{{z_1}^{1+2\epsilon}(z+z_2)},
\end{eqnarray}
\end{subequations}
%--------------------------------------
where 
%--------------------------------------
\begin{equation}
\int_{{z_1}{z_2}}\!\!\!\!F({z_1},{z_2})\,\,=
\int_0^1\!d{z_1}\int_0^1\!d{z_2}
\,\,
F({z_1},{z_2})
\,
\delta(1-z-{z_1}-{z_2}).
\end{equation}
%--------------------------------------
By making use of the identity $z_2/(z+z_2)= 1-z/(z+z_2)$, 
we derive the relations
%--------------------------------------
\begin{subequations}
\label{integral-recursion}
\begin{eqnarray}
A_n&=&B_{n-1}-zA_{n-1},\\
D_n&=&C_{n-1}-zD_{n-1}.
\end{eqnarray}
\end{subequations}
%--------------------------------------
Applying the recursion relations (\ref{integral-recursion}) repeatedly,
we can reduce all of the integrals to the forms $A_0$, $B_n$ and $C_n$,
with $1\leq n\leq 4$. It turns out that the coefficient of $D_0$
vanishes. 
The coefficient of $A_0$ is of order $\epsilon^0$, while the
coefficients of the $B_n$ are of order $\epsilon^{-1}$ or $\epsilon^0$.
Therefore, we evaluate $A_0$ only through order $\epsilon^0$, and we
evaluate the $B_n$ only through order $\epsilon$.
The expressions for
$A_0$, $B_n$ and $C_n$ can be obtained conveniently by 
making use of the identity
%--------------------------------------
\begin{equation}
\label{eq:z11p2e}
\frac{1}{z_1^{1+2\epsilon}}=
-\frac{(1-z)^{-2\epsilon}}{2\epsilon}\,\delta(z_1)
+\left[\frac{1}{z_1^{1+2\epsilon}}\right]_{1-z},
\end{equation}
%--------------------------------------
which applies when the domain of integration is $0\le z_1\le 1-z$.
The distribution in the second term of Eq.~(\ref{eq:z11p2e})
is defined by
%--------------------------------------
\begin{equation}
\int_0^{1-z}\!\!dz_1\, f(z_1)
\left[\frac{1}{z_1^{1+2\epsilon}}\right]_{1-z} =
\int_0^{1-z}\!\!dz_1\, 
\frac{f(z_1)-f(0)}{z_1^{1+2\epsilon}}.
\end{equation}
%--------------------------------------
A straightforward evaluation of the integrals then gives
%--------------------------------------
\begin{subequations}
\begin{eqnarray}
A_0
&=&
-\frac{1}{2\epsilon(1-z)^{3\epsilon}}
-\log z
+O(\epsilon),\\
B_n
&=&
-\frac{(1-z)^{n-3\epsilon}}{2\epsilon}
+
(1-z)^{n-3\epsilon}
\,
\big[
X_n+Y_n+Z_n+O(\epsilon^{2})
\big]
,\\
\label{app-c_n}
C_n
&=&
(1-z)^{n-4\epsilon}
\left\{
-
\frac{1}{2\epsilon}
+
\frac{1}{2\epsilon}
\left[
1-
\frac{\Gamma(n+1-2\epsilon)\Gamma(1-2\epsilon)}
{\Gamma(n+1-4\epsilon)}
\right]
\right\},
\end{eqnarray}
\end{subequations}
%--------------------------------------
where 
%--------------------------------------
\begin{subequations}
\begin{eqnarray}
X_n&=&\int_0^1 \! dt\, \, 
\frac{(1-t)^n-1}{t}
=-\sum_{k=1}^{n}\frac{1}{k},
\\
Y_n&=&-\epsilon\int_0^1 \! dt\, \, 
\frac{(1-t)^n}{t}
\big\{\log(1-t)+\log[1-(1-z) t]\,\big\},
\\
Z_n&=&-2\epsilon\int_0^1 \! dt\, \, 
\frac{(1-t)^n-1}{t}\log t=
-2\epsilon
\sum_{k=1}^{n}\frac{1}{k}
\sum_{\ell=1}^{k}
\frac{1}{\ell}.
\end{eqnarray}
\end{subequations}
%--------------------------------------
In Eq.~(\ref{app-c_n}), the first term in the braces is the pole contribution, 
and the remainder is finite.
The results for $Y_n$ for $0\leq n\leq 4$ are
%--------------------------------------
\begin{subequations}
\begin{eqnarray}
Y_0&=&
\epsilon \Bigg[\frac{\pi^2}{6} + {\rm Li}_2(1-z)\Bigg],
\\
Y_1&=&
\epsilon \Bigg[\frac{\pi^2}{6}+ {\rm Li}_2(1-z)  
-2
-\frac{z}{1-z}\log z  
\Bigg],
\\
Y_2&=&
\epsilon
       \Bigg[
    \frac{\pi^2}{6}+{\rm Li}_2(1-z) 
   -\frac{5-6z}{2(1-z)} -\frac{z(2-3z)\log z }{2(1-z)^2}
       \Bigg],
\\
Y_3&=&
\epsilon \Bigg[
\frac{\pi^2}{6}+{\rm Li}_2(1-z) 
-\frac{49  - 110 z + 67 z^2 }{18(1-z)^2}
-\frac{z(6 -15 z + 11 z^2)\log z }{6(1-z)^3}
\Bigg],\phantom{xxx}
\\
Y_4&=&
\epsilon 
\Bigg[
\frac{\pi^2}{6}+{\rm Li}_2(1-z) 
-\frac{205-669 z+756 z^2-310 z^3}{72(1-z)^3}
\nonumber\\
&&\quad
-\frac{z(12-42 z+52 z^2-25 z^3)\log z}{12(1-z)^4}
\Bigg].\phantom{xxxxxx}
\end{eqnarray}
\end{subequations}
%%%%%%%%%%%%%%%%%%%%%%%%%%%%%%%%%%%%%%%%%%%%%%%%%%%%%%%%%%%%%%%%%%%%%%
%%%%%%%%%%%%%%%%%%%%%%%%%%%%%%%%%%%%%%%%%%%%%%%%%%%%%%%%%%%%%%%%%%%%%%
\section{Parametrizations of the short-distance
coefficients\label{app:param}}
%%%%%%%%%%%%%%%%%%%%%%%%%%%%%%%%%%%%%%%%%%%%%%%%%%%%%%%%%%%%%%%%%%%%%%
%%%%%%%%%%%%%%%%%%%%%%%%%%%%%%%%%%%%%%%%%%%%%%%%%%%%%%%%%%%%%%%%%%%%%%
In this appendix, we give parametrizations for the 
short-distance coefficients
$d_0[g\to Q\bar{Q}({}^3S_1^{[1]})]^{(3)}(z)$,
$d_2[g\to Q\bar{Q}({}^3S_1^{[1]})]^{(3)}(z)$, and
$d_4[g\to Q\bar{Q}({}^3S_1^{[1]})]^{\rm finite}(z)$, which are defined
in Eqs.~(\ref{D0-J/psi}), (\ref{D2-J/psi}), and (\ref{d4-finite}),
respectively.

We observe that $d_0[g\to Q\bar{Q}({}^3S_1^{[1]})]^{(3)}(z)$,
$d_2[g\to Q\bar{Q}({}^3S_1^{[1]})]^{(3)}(z)$, and
$d_4[g\to Q\bar{Q}$
$({}^3S_1^{[1]})]^{\rm finite}(z)$ are continuous 
functions of $z$ over the whole range $0\le z\le 1$ and that they are
not analytic at the endpoints $z=0$ and $1$. One can prove that $d_n(0)=0$ for 
$n=0$, 2, and 4.  In addition, $d_n(1)=\alpha_s^3 b_n/m^{3+n}$ is finite
and calculable analytically. The values of $b_n$ for $n=0$, 2, and 4 are 
%--------------------------------------
\begin{subequations}
\begin{eqnarray}
b_{0}&=&0,
\\
b_{2}&=&\frac{22\pi^2-15}{4374\pi},
\\
b_{4}&=&-\frac{2922 \pi^2-2485}{229635 \pi}.
\end{eqnarray}
\end{subequations}
%--------------------------------------

Using this information, we parametrize 
$d_0[g\to Q\bar{Q}({}^3S_1^{[1]})]^{(3)}(z)$,
$d_2[g\to Q\bar{Q}({}^3S_1^{[1]})]^{(3)}(z)$, 
and
$d_4[g\to Q\bar{Q}({}^3S_1^{[1]})]^{\rm finite}(z)$
as follows:
%--------------------------------------
\begin{eqnarray}
\label{eq:param}
d_n^{\rm fit}(z)=
\frac{\alpha_s^3}{m^{3+n}}&\Bigg[&
b_n z+
\log (1-z)\sum_{k=1}^{n_a}\alpha_{nk} (1-z)^k
+\log^2(1-z)
\sum_{k=1}^{n_b} \beta_{nk} (1-z)^k
\nonumber\\
&&
+
\log z\sum_{k=1}^{n_c} \mu_{nk} z^k
+
\log^2 z
\sum_{k=1}^{n_d} \nu_{nk} z^k
+
\sum_{k_1=1}^{n_1}
\sum_{k_2=1}^{n_2}
 \omega_{nk_1k_2} z^{k_1}(1-z)^{k_2}\Bigg].\nonumber\\
\end{eqnarray}
In order to fix the parameters in Eq.~(\ref{eq:param}), we have computed 
$d_0[g\to Q\bar{Q}({}^3S_1^{[1]})]^{(3)}(z)$,
$d_2[g\to Q\bar{Q}({}^3S_1^{[1]})]^{(3)}(z)$, and
$d_4[g\to Q\bar{Q}({}^3S_1^{[1]})]^{\rm finite}(z)$ by numerical 
integration for $100$ points in each of the ranges  $z=0$ to $z=10^{-2}$,
$z=10^{-2}$ to $z=1-10^{-2}$, and $z=1-10^{-2}$ to $z=1$. 
We have carried out the integrations in two ways: (1) using the change of
variables that was proposed in
Refs.~\cite{Braaten:1993rw,Braaten:1995cj} and is described in Appendix~B
of Ref.~\cite{Bodwin:2003wh} and (2) using the change of variables that
is given in Eq.~\eqref{u1-u2}. At each value of $z$, the numerical
results from the two methods of integration agree to better than 
$\Delta^{\rm int}_n=r_n d_n(z)$, where 
             $(r_0,r_2,r_4) = (7.8 \times 10^{-4}, 1.7 \times 10^{-3}, 1.8 \times 10^{-3})$.
%============================================================================
\begin{table}[t]
\caption{\label{tab:dz}%
Results of fitting the parametrization in Eq.~(\ref{eq:param}) to
numerical values for the color-singlet short-distance coefficients
$d_0[g \to Q\bar{Q}({}^3S_1^{[1]})]^{(3)}(z)$,
$d_2[g \to Q\bar{Q}({}^3S_1^{[1]})]^{(3)}(z)$, and
$d_4[g \to Q\bar{Q}({}^3S_1^{[1]})]^{\rm finite}(z)$,
which are defined in Eqs.~(\ref{D0-J/psi}), (\ref{D2-J/psi}), and
(\ref{d4-finite}), respectively. The quantity $\Delta_{\rm max}^{\rm
fit}$ is described in the text.}
\begin{center}
\begin{tabular}{|cccccc|}
\hline
Parameters&
{$d_0(z)$}&\phantom{xxx}&
{$d_2(z)$}&\phantom{xxx}&
{$d_4(z)^{\rm finite}$}
\\
\hline
$b_{\phantom{11}}$&$
\phantom{-}0\phantom{11111111111}$&&
$\phantom{-}1.4710\times 10^{-2}$&&
$-3.6531\times 10^{-2}$\\
\hline
$\alpha_{1\phantom{1}}$&
$-4.9866\times 10^{-3}$&&
$-1.8127\times 10^{-2}$&&
$\phantom{-}5.2157\times 10^{-2}$\\
$\alpha_{2\phantom{1}}$&
$\phantom{-}9.8448\times 10^{-3}$&&
&&
$\phantom{-}2.4588\times 10^{-3}$\\
$\alpha_{3\phantom{1}}$&
$\phantom{-}1.9512\times 10^{-2}$&&
&&\\
\hline
$\beta_{1\phantom{1}}$&
$-5.1697\times 10^{-6}$&&
$-1.2282\times 10^{-2}$&&
$\phantom{-}3.6020\times 10^{-2}$\\
$\beta_{2\phantom{1}}$&
$\phantom{-}1.0462\times 10^{-2}$&&
&&
\\
\hline
$\mu_{1\phantom{1}}$&
$-1.8921\times 10^{-3}$&&
$\phantom{-}4.1069\times 10^{-3}$&&
$-7.3565\times 10^{-3}$\\
$\mu_{2\phantom{1}}$&&&$\phantom{-}1.8341\times 10^{-2}$&&
$-2.5387\times 10^{-2}$\\
$\mu_{3\phantom{1}}$&&&&&$-9.3477\times 10^{-3}$\\
$\mu_{4\phantom{1}}$&&&&&$-3.6926\times 10^{-3}$\\
\hline
$\nu_{1\phantom{1}}$&
$\phantom{-}1.2154\times 10^{-3}$&&
$-1.3653\times 10^{-3}$&&
$\phantom{-}1.3839\times 10^{-3}$\\
$\nu_{2\phantom{1}}$&$\phantom{-}1.3039\times 10^{-3}$&&&&\\
$\nu_{3\phantom{1}}$&$-2.7246\times 10^{-3}$&&&&\\
$\nu_{4\phantom{1}}$&$-1.4814\times 10^{-3}$&&&&\\
\hline
$\omega_{11}$&$-1.6910\times 10^{-2}$&&
$-2.1832\times 10^{-2}$&&
$\phantom{-}7.7264\times 10^{-2}$\\
$\omega_{12}$&
$\phantom{-}3.8110\times 10^{-2}$&&
$\phantom{-}4.1531\times 10^{-3}$&&
\\
$\omega_{13}$&&&$\phantom{-}8.4949\times 10^{-4}$&&
\\
$\omega_{14}$&&&$-3.8207\times 10^{-3}$&&\\
\hline
$\Delta_{\rm max}^{\rm fit}\times m^{3+n}/\alpha_s^3$&
\,$3.26\times 10^{-8}\phantom{1}$&&
\,$1.12\times 10^{-6}\phantom{1}$&&
\,$2.05\times 10^{-6}\phantom{1}$\\
\hline
\end{tabular}
\end{center}
\end{table}
%============================================================================

We fit the parametrizations to the numerical integration results, using $\chi^2=\sum_i
[d(z_i)-d^{\rm fit}(z_i)]^2/[\sigma(z_i)]^2$ as the criterion for goodness
of fit, where $\sigma(z_i)$ is the error in the numerical
integration that is given by the \textsc{VEGAS} integration program
\cite{Lepage:1977sw}. In cases in which it is possible to reduce the
number of parameters in the fit without significantly affecting 
$\chi^2$, we have done so.

In Table~\ref{tab:dz}, we show the
results of this fitting procedure, along with the value for each fit of
$\Delta_{\rm max}^{\rm fit}=\textrm{Max}|d(z_i)-d^{\rm fit}(z_i)|$.
In these fits, $\chi^2$ per
degree of freedom is much larger than one because the error in the fit
is generally considerably larger than the errors in the numerical
integrations that are given by \textsc{VEGAS}. A conservative estimate of the
overall error in a parametrization at each value of $z$ is
$\sqrt{[\Delta_{\rm max}^{\rm fit}]^2+[\Delta_n^{\rm int}]^2}$.

%%%%%%%%%%%%%%%%%%%%%%%%%%%%%%%%%%%%%%%%%%%%%%%%%%%%%%%%%%%%%%%%%%%%%%%%%%%%%
%%%%%%%%%%%%%%%%%%%%%%%%%%%%%%%%%%%%%%%%%%%%%%%%%%%%%%%%%%%%%%%%%%%%%%%%%%%%%

\end{document}